\documentclass[reqno]{amsproc}
\makeatletter
\RequirePackage{amsthm}[1996/09/24]
\def\@swapped#1#2{#2%
  \@ifnotempty{#1}{\@addpunct{.}\quad#1\unskip}}
\def\thmhead@plain#1#2#3{%
  \thmname{#1}\thmnumber{\@ifnotempty{#1}{ }\@upn{#2}}%
  \thmnote{ \textmd{\upshape(#3)}}}
\def\swappedhead@plain#1#2#3{%
  \thmnumber{\@upn{#2}}\thmname{\@ifnotempty{#2}{. }#1}%
  \thmnote{ \textmd{\upshape(#3)}}}
\def\th@plain{%
  \let\thm@indent\indent
  \thm@headfont{\bfseries}
  \let\thmhead\thmhead@plain \let\swappedhead\swappedhead@plain
  \thm@preskip.5\baselineskip\@plus.2\baselineskip
                                    \@minus.2\baselineskip
  \thm@postskip\thm@preskip
  \itshape
}
\def\th@definition{%
  \let\thm@indent\indent
  \thm@headfont{\bfseries}
  \let\thmhead\thmhead@plain \let\swappedhead\swappedhead@plain
  \thm@preskip.5\baselineskip\@plus.2\baselineskip
                                    \@minus.2\baselineskip
  \thm@postskip\thm@preskip
  \upshape
}
\def\th@remark{%
  \let\thm@indent\indent
  \thm@headfont{\bfseries}
  \let\thmhead\thmhead@plain \let\swappedhead\swappedhead@plain
  \thm@preskip.5\baselineskip\@plus.2\baselineskip
                                    \@minus.2\baselineskip
  \thm@postskip\thm@preskip
  \upshape
}
\makeatother
\makeatletter

\newcounter{numcount}
\newcommand{\labelnummer}{\mbox{(\roman{numcount})}}%

\newenvironment{indentnummer}%
    {\let\curlabelspeicher\@currentlabel%
     \begin{list}{\labelnummer}{\usecounter{numcount}%
                  \topsep1ex\partopsep2ex\parsep0pt\itemsep1.5ex\@plus1\p@%
                  \labelwidth3em\itemindent0em\labelsep1em%
                  \leftmargin3em}%
     \let\saveitem\item%
     \def\item{\saveitem%
               \def\@currentlabel{\curlabelspeicher\labelnummer}%
               \let\label\bemlabel}}%
   {\end{list}}%

\newenvironment{indentnummer*}%
    {\begin{list}{\labelnummer}{\usecounter{numcount}%
                  \topsep1ex\partopsep2ex\parsep0pt\itemsep1ex
                  \labelwidth2.5em\itemindent0em\labelsep1em%
                  \leftmargin3.5em
                  }}%
   {\end{list}}%

\newenvironment{nummer}%
    {\let\curlabelspeicher\@currentlabel%
     \begin{list}{\labelnummer}{\usecounter{numcount}\leftmargin0em%
                  \topsep1ex\partopsep2ex\parsep0pt\itemsep0.5ex
                  \labelwidth2.5em\itemindent3.5em\labelsep1em}%
     \let\saveitem\item%
     \def\item{\saveitem%
               \def\@currentlabel{\curlabelspeicher\labelnummer}%
               \let\label\bemlabel}}%
   {\end{list}}%

\def\itemref#1{\expandafter\@setref\csname r@#1item\endcsname\@firstoftwo{#1}}%

\def\bemlabel#1{\@bsphack%
  \protected@write\@auxout{}%
         {\string\newlabel{#1}{{\@currentlabel}{\thepage}}}%
  \ifmmode\else%
  \protected@write\@auxout{}%
         {\string\newlabel{#1item}{{\labelnummer}{\thepage}}}%
  \fi%
  \@esphack}%
\makeatother

\usepackage{bbm}
\usepackage{times}
\usepackage{amsthm}

\newtheorem{theorem}{Theorem}[section]
\newtheorem{proposition}[theorem]{Proposition}
\newtheorem{lemma}[theorem]{Lemma}

\theoremstyle{definition}

\newtheorem{remark}[theorem]{Remark}
\newtheorem{remarks}[theorem]{Remarks}
\newtheorem{assumption}[theorem]{Assumption}

\def\infspec{\mathop{\hbox{\rm inf spec}}}

\begin{document}
\bibliographystyle{alpha}

\title[Lifshits tails caused by anisotropic decay]{Lifshits tails caused by anisotropic decay:\\ the emergence of a quantum-classical regime}

\author{Werner Kirsch}
\email{werner.kirsch@mathphys.ruhr-uni-bochum.de}
\address{%
Institut f\"ur Mathematik,
          Ruhr-Universit\"at Bochum und SFB TR 12, 44780 Bochum, Germany
}%
\author{Simone Warzel}%
\email{swarzel@princeton.edu}
\address{%
\textit{Present address:}
Princeton University, Deptartment of Physics, Jadwin Hall,
Princeton, NJ 08544, USA }
\address{%
\textit{On leave from:} \\
 Institut f\"ur Theoretische Physik, Universit\"at Erlangen-N\"urnberg, Staudtstr.\ 7, 91058 Erlangen,
Germany
}%

\begin{abstract}
  We investigate Lifshits-tail behaviour of the integrated density of states for a wide class of Schr\"odinger operators with
  positive random potentials. The setting includes alloy-type
  and Poissonian random potentials.
  The considered (single-site) impurity
  potentials $f: \mathbbm{R}^d \to [0, \infty[ $ decay at infinity in an anisotropic way, for example,
  $f(x_1,x_2)\sim (|x_1|^{\alpha_1}+|x_2|^{\alpha_2})^{-1}$ as $ |(x_1,x_2)| \to \infty $.
  As is expected from the isotropic situation, there is a so-called quantum regime with Lifshits exponent $ d/2 $
  if both $\alpha_1$ and $\alpha_2$ are big enough, and there is a so-called classical regime with Lifshits exponent depending on
  $\alpha_1$ and $\alpha_2$ if both are small. In addition to this we find two new regimes where the Lifshits exponent exhibits a mixture of
  quantum and classical behaviour. Moreover, the transition lines between these regimes depend in a
  nontrivial way on $ \alpha_1$ and $\alpha_2$ simultaneously.
\end{abstract}


\maketitle
\begin{center}
Dedicated to the memory of G. A. Mezincescu ( 1943 -- 2001 ).
\end{center}
\setcounter{tocdepth}{2}
\tableofcontents
%
%
\section{Introduction}
The integrated density of states $ N : \mathbbm{R} \to [0, \infty[ $ is an important basic quantity in the theory of
disordered electronic systems \cite{Kir89,CaLa90,Lan91,PaFi92,Sto01,LeMuWa03,Ves03}.
Roughly speaking, $ N(E) $ describes the number of energy levels below a given energy $ E $ per unit volume
(see (\ref{eq:defN}) below for a precise definition).
A characteristic feature of disordered systems is the behaviour of $ N $ near band edges. It was first studied by
Lifshits~\cite{Lif63}.
He gave convincing physical arguments that the polynomial decrease
\begin{equation}
  \log N(E) \sim  \log \; (E- E_0)^{\frac{d}{2}} \qquad \mbox{as} \quad E \downarrow E_0
\end{equation}
known as van-Hove singularity (see \cite{KirSim87} for a rigorous proof) near a band edge $ E_0 $ of an ideal periodic system in $d$ space dimensions
is replaced by an exponential decrease in a disordered system.
In his honour, this decrease is known as Lifshits singularity or Lifshits tail and typically given by
\begin{equation} \label{eq:LifS}
 \log N(E) \sim \log  e^{-C \; (E-E_0)^{-\eta}} \qquad \mbox{as} \quad E \downarrow E_0
\end{equation}
where $ \eta > 0 $ is called the Lifshits exponent and $ C > 0 $ is some constant.

The first rigorous proof \cite{DonVar75}
(see \cite{Nak77}) of Lifshits tails (in the sense that (\ref{eq:LifS}) holds)
concerns the bottom $ E_0 $ of the energy spectrum of a continuum model involving a Poissonian random potential
\begin{equation}\label{eq:defpois}
  V_\omega(x):=\sum_j\:f(x-\xi_{\omega,j}),
\end{equation}
where $\xi_{\omega,j} \in \mathbbm{R}^d$ are Poisson distributed points and $f: \mathbbm{R}^d \to [0, \infty[$ is a non-negative impurity potential.
Donsker and Varadhan \cite{DonVar75}
particularly showed that the Lifshits exponent is universally given by $ \eta = d/2 $ in case
\begin{equation}\label{eq:fDV}
  0 \leq f(x)\,\le\,f_0\,(1+|x|)^{-\alpha} \quad \mbox{with some $\alpha > d+2$  and some $ f_0 > 0 $.}
\end{equation}
It was Pastur \cite{Pas77} who proved that the Lifshits
exponent changes to $ \eta = d/(\alpha- d) $ if
\begin{equation}\label{eq:fPas}
  f_u \,(1+|x|)^{-\alpha} \leq f(x) \leq f_0 \,(1+|x|)^{-\alpha}  \quad \begin{array}{c} \mbox{with some  $d<\alpha<d+2$}\\
                                                                                        \mbox{and some $ f_u$, $ f_0 > 0 $.}
                                                                                        \end{array}
\end{equation}
This change from a universal Lifshits exponent to a non-universal one, which depends on the decay exponent $
\alpha $ of $ f $, may be heuristically explained in terms of a competition of the kinetic and the potential
energy of the underlying one-particle Schr{\"o}dinger operator. In the first case ($\eta=d/2$) the quantum
mechanical kinetic energy has a crucial influence on the (first order) asymptotics of $N$. The Lifshits tail is
then said to have a \emph{quantum} character. In the other case it is said to have a \emph{classical} character
since then the (classical) potential energy determines the asymptotics of $N$. For details, see for example
\cite{Lan91,PaFi92,LeWa03}.

Analogous results have been obtained for other random potentials. For example, in case of an alloy-type random potential
\begin{equation}\label{eq:defRP}
  V_\omega(x) := \sum_{j \in \mathbbm{Z}^d} q_{\omega,j} \, f(x-j)
\end{equation}
which is given in terms of independent identically distributed random variables $q_{\omega,j}$ and an impurity
potential $f: \mathbbm{R}^d \to [0, \infty[ $, the Lifshits tails at the lowest band edge $ E_0 $ have been
investigated by \cite{KirMar83a,KirSim86,Mez87}. Similarly to the Poissonian case the authors of
\cite{KirSim86,Mez87} consider $ f $ as in (\ref{eq:fDV}) and (\ref{eq:fPas}) and detect a quantum and a classical
regime for which the Lifshits exponent equals
\begin{equation}\label{eq:lifexp}
  \eta  =  \left\{ \begin{array}{c@{\quad\mbox{in case}\,}l}
      \frac{d}{2} &  (\ref{eq:fDV}): \quad d + 2 < \alpha \\[1ex]
       \frac{d}{\alpha - d} & (\ref{eq:fPas}): \quad d < \alpha < d +2
      \end{array}
      \right\}  = \max\left\{ \frac{d}{2}, \frac{d/\alpha}{1-d/\alpha} \right\}
\end{equation}
In fact they do not obtain the asymptotics (\ref{eq:LifS}) on a logarithmic scale but only double-logarithmic asymptotics (confer (\ref{eq:Hauptresultat}) below).
(See also \cite{Sto99} for an alternative proof of this double-logarithmic asymptotics in case of alloy-type and Poissonian random potentials.)

Our main point is to generalise these results on the Lifshits exponent to impurity potentials $ f $ that decay in
an \emph{anisotropic} way at infinity (confer (\ref{eq:defV}) below). In addition we are able to handle a wide
class of random potentials given in terms of random Borel measures which include among further interesting
examples both the case of alloy-type potentials and Poisson potential. Thus the \emph{same} proof works for
these two most important cases.

In our opinion it is interesting to explore the transition between quantum  and classical Lifshits behaviour in
such models from both a mathematical and a physical point of view. The interesting cases are those for which $f$
decays fast enough in some directions to ensure a quantum character while it decays slowly in the other direction
so that the expected character there is the classical one. In the following we give a complete picture of the
classical and the quantum regime of the integrated density of states as well as of the emerging mixed
\emph{quantum-classical} regime. We found it remarkable that the  borderline between the quantum and classical
behaviour caused by the decay of $ f $ in a certain direction is not determined by the corresponding decay
exponent of these directions alone, but depends also in a nontrivial way on the decay in the other directions.

A second motivation for this paper came from investigations of the Lifshits tails
in a constant magnetic field in three space dimensions \cite{War01,HuKiWa03,LeWa03}.
In contrast to the two-dimensional situation \cite{BHKL95,Erd98,HuLeWa99,HuLeWa00,Erd01,War01},
the magnetic field introduces an anisotropy in $ \mathbbm{R}^3 $, such that it is quite natural to look at $ f $ which are
anisotropic as well.  In fact, in the three-dimensional magnetic case a quantum-classical regime has already been shown to occur
for certain $ f $ with isotropic decay \cite{War01,LeWa03}.
The present paper will contribute to a better understanding of these results.

The results mentioned above as well as the results in this paper concern Lifshits tails at the bottom of the spectrum.
In accordance with Lifshits' heuristics, the integrated
density of states should behave in a similar way at other edges of the spectrum. Such internal Lifshits tails were proven in
\cite{Mez86,Sim87,Mez93,Klo99,KlWo02,Klo02}.\\

\noindent
{\bf Acknowledgement:} We are grateful to Hajo Leschke for helpful remarks.
This work was partially supported by the DFG within the SFB TR 12.

\section{Basic quantities and main result}
\subsection{Random potentials}\label{sec:ass}
We consider random potentials
\begin{equation}\label{eq:defV}
 V: \Omega \times \mathbbm{R}^d \to [0, \infty [, \quad  (\omega, x) \mapsto V_\omega(x) := \int_{\mathbbm{R}^d} f(x-y) \, \mu_\omega(dy),
\end{equation}
which are given
in terms of a random Borel measure $ \mu:  \Omega \to \mathcal{M}(\mathbbm{R}^d) $, $ \omega \mapsto \mu_\omega $,
and an impurity potential $ f : \mathbbm{R}^d \to [0, \infty[ $.
We recall from \cite{Kal83,SKM87,DaVeJo88} that a random Borel measure is a measurable mapping
from a probability space $ \left( \Omega , \mathcal{A}, \mathbbm{P} \right) $
into the set of Borel measures $ \big(\mathcal{M}(\mathbbm{R}^d), \mathcal{B}(\mathcal{M}) \big)$, that is, the set of positive, locally-finite measures
on $ \mathbbm{R}^d $. Here
$ \mathcal{B}(\mathcal{M}) $ denotes the Borel $ \sigma $-algebra of $ \mathcal{M}(\mathbbm{R}^d) $, that is, the smallest $ \sigma $-algebra rendering the mappings
$ \mathcal{M}(\mathbbm{R}^d) \ni \nu \mapsto \nu(\Lambda) $ measurable for all bounded Borel sets $ \Lambda \in \mathcal{B}(\mathbbm{R}^d) $.

The following assumptions on $ \mu $ are supposed to be valid throughout the paper.
\begin{assumption}\label{ass:q}
The random Borel measure $ \mu:  \Omega \to \mathcal{M}(\mathbbm{R}^d)  $, $ \omega  \mapsto \mu_\omega $ is
defined on some complete probability space $ \left( \Omega , \mathcal{A}, \mathbbm{P} \right) $.
We suppose that:
\begin{indentnummer}
\item\label{stat}
  $ \mu $ is $ \mathbbm{Z}^d $-stationary.
\item\label{indep}
  there exists a partition of $ \mathbbm{R}^d = \bigcup_{j \in \mathbbm{Z}^d} \Lambda_j $ into disjoint unit cubes $  \Lambda_j = \Lambda_0 + j$
  centred at the sites of the lattice $ \mathbbm{Z}^d $
  such that the random variables $ \big( \mu(\Lambda^{(j)} \big) _{j \in J} $ are stochastically independent for any finite collection $ J \subset \mathbbm{Z}^d $ of
  Borel sets $ \Lambda^{(j)} \subset \Lambda_j $.
\item\label{intensitymeas}
  the intensity measure $ \overline{\mu}:\mathcal{B}(\mathbbm{R}^d) \to [0,\infty[ $, which is given by
  \begin{equation}
     \overline{\mu}(\Lambda) := \mathbbm{E}\big[  \mu(\Lambda) \big]
  \end{equation}
  in terms of the probabilistic expectation $ \mathbbm{E}[ \cdot ]:= \int_\Omega (\cdot) \, \mathbbm{P}(d\omega) $, is a Borel measure which does not vanish identically
  $ \overline{\mu} \neq 0 $.
\item\label{prob}
  there is some constant $ \kappa > 0 $ such that $ \mathbbm{P}\left\{ \omega \in \Omega : \, \mu_\omega(\Lambda_0) \in [0, \varepsilon[ \right\} \geq \varepsilon^\kappa $
  for small enough $ \varepsilon > 0 $.
\end{indentnummer}

\end{assumption}

\begin{remark}
    Assumption~\ref{stat} implies that the intensity measure $ \overline{\mu} $ is $ \mathbbm{Z}^d $-periodic.
    Assumption~\ref{intensitymeas} is thus equivalent to the existence of the first moment
    $ \mathbbm{E}\left[ \mu(\Lambda_0) \right] < \infty $ of the random variable $ \mu(\Lambda_0): \omega \mapsto  \mu_\omega(\Lambda_0) $.
    Moreover, we emphasis that the unit cubes $ (\Lambda_j) $ introduced in Assumption~\ref{indep} are neither open nor closed.
\end{remark}

We recall from \cite{Kal83,SKM87,DaVeJo88} that $  \mathbbm{Z}^d $-stationarity of $ \mu $ requires
the group $ (T_j)_{j \in \mathbbm{Z}^d} $ of lattice translations, which is defined on $ \mathcal{M}(\mathbbm{R}^d ) $
by $ (T_j \nu)(\Lambda) := \nu(\Lambda + j) $ for
all $ \Lambda \in \mathcal{B}(\mathbbm{R}^d) $ and all $ j \in \mathbbm{Z}^d $, to be probability preserving in the sense that
\begin{equation}
  \mathcal{P}\left\{ T_j M \right\} =  \mathcal{P}\left\{ M \right\}
\end{equation}
for all $  M \in \mathcal{B}(\mathcal{M}) $ and all $ j \in \mathbbm{Z}^d $. Here we have introduced the notation
$ \mathcal{P}\left\{ M \right\} := \mathbbm{P}\left\{ \omega \in \Omega \, : \, \mu_\omega \in M \right\} $ for
the induced probability measure on $ \big( \mathcal{M}(\mathbbm{R}^d) , \mathcal{B}(\mathcal{M}) \big) $. To
ensure the ($ \mathbbm{Z}^d$-)ergodicity of the random potential $ V $, it is useful to know that under the assumptions made above,
$ (T_j) $ is a
group of mixing (hence ergodic) transformations on the probability space $\big( \mathcal{M}(\mathbbm{R}^d) ,
\mathcal{B}(\mathcal{M}), \mathcal{P}\big) $.

\begin{lemma}\label{lemma:mix}
Assumption~\ref{stat} and \ref{indep} imply that $ \mu $ is mixing in the sense that
\begin{equation}\label{eq:mix}
  \lim_{|j| \to \infty } \mathcal{P}\left\{ T_j M \cap M' \right\} = \mathcal{P}\left\{ M \right\} \mathcal{P}\left\{ M' \right\}
\end{equation}
for all $  M, M' \in \mathcal{B}(\mathcal{M}) $.
\end{lemma}
\begin{proof} See Appendix~\ref{app:mixing}.
\end{proof}

The considered impurity potentials $ f : \mathbbm{R}^d \to [0, \infty[ $ comprise a large class of functions with anisotropic decay.
More precisely, we decompose the configuration
space $ \mathbbm{R}^d = \mathbbm{R}^{d_1} \times \dots \times \mathbbm{R}^{d_m} $
into $ m \in \mathbbm{N} $ subspaces with dimensions $ d_1, \dots, d_m \in \mathbbm{N} $.
Accordingly, we will write $ x = (x_1, \dots , x_m) \in \mathbbm{R}^d $, where $ x_k \in \mathbbm{R}^{d_k} $ and $ k \in \{1, \dots , m\} $.
Denoting by $ |x_k | := \max_{i \in \{1, \dots, d_k\}} | (x_k)_i | $ the maximum norm on $ \mathbbm{R}^{d_k} $,
our precise assumptions on $ f $ are as follows.
\begin{assumption}\label{ass:f}
      The impurity potential $ f : \mathbbm{R}^d \to [0,\infty [ $ is positive, strictly positive on some non-empty open set and satisfies:
      \begin{indentnummer}
        \item
          the Birman-Solomyak condition
          $ \sum_{j \in \mathbbm{Z}^d} \big(\int_{\Lambda_0} |f(x-j)|^p dx \big)^{1/p} < \infty $ with $ p = 2 $ if $ d \in \{1,2,3\} $ and $ p > d/2 $ if $ d \geq 4 $.
        \item\label{pint}
          there exist constants $ \alpha_1, \dots , \alpha_m \in [0, \infty] $ and $ 0 < f_u$, $ f_0 < \infty $ such that
          \begin{equation}\label{eq:assf}
            \frac{f_u}{\sum_{k=1}^{m} |x_k|^{\alpha_k}} \leq \int_{\Lambda_0} \! f(y-x) \, dy, \qquad
            f(x) \leq \frac{f_0}{\sum_{k=1}^{m} |x_k|^{\alpha_k}}
          \end{equation}
          for all $  x = (x_1, \dots , x_m) \in \mathbbm{R}^d $ with large enough values of their maximum norm
       $ | x | = \max\{ |x_1|, \dots , |x_m| \} $.
      \end{indentnummer}
\end{assumption}
\begin{remark}\label{rem:infty}    In order to simultaneously treat the case $ \alpha_k = \infty $  for some (or all) $ k \in \{1, \dots , m\} $, we adopt the conventions
    $ | x_k |^\infty := \infty $ for $ | x_k | > 0 $ and $ 1/ \infty := 0 $.
    An example for such a situation is given by $ f $ with compact support in the $ x_k $-direction.
\end{remark}

\subsection{Examples}
The setting in Subsection~\ref{sec:ass} covers a huge class of random potentials which are widely
encountered in the literature on random Schr{\"o}dinger operators \cite{Kir89,CaLa90,PaFi92,Sto01}.
In this Subsection we list prominent examples, some of which have already been (informally) introduced in the Introduction.\\

From the physical point of view, it natural to consider integer-valued random Borel measures $ \nu = \sum_j k_{j} \delta_{x_{j}} $,
also known as point processes \cite{DaVeJo88}.
Here each $ k_j $ is an integer-valued random variable and the distinct points $ (x_j) $ indexing the atoms, equivalently the Dirac measure $ \delta $, form
a  countable (random) set with at most finitely many $ x_j $ in any bounded Borel set.
In fact, interpreting $ (x_j) $ as the (random) positions of impurities in a disordered solid
justifies the name 'impurity potential' for $ f $ in (\ref{eq:defV}).

Two examples of point processes satisfying Assumptions~\ref{stat}--\ref{intensitymeas} are:
\begin{indentnummer*}
  \item[(P)]\label{genPoiss} the \emph{generalised Poisson measure} $ \nu = \sum_j \delta_{\xi_{j}} $ with some non-zero $ \mathbbm{Z}^d $-periodic
    Borel intensity measure $ \overline{\nu} $.
    The Poisson measure is uniquely characterised by requiring that the random variables
    $ \nu(\Lambda^{(1)}), \dots , \nu(\Lambda^{(n)}) $ are stochastically independent for any collection of disjoint
    Borel sets $ \Lambda^{(1)}, \dots \Lambda^{(n)} \in \mathcal{B}(\mathbbm{R}^d) $ and that each $ \nu(\Lambda) $ is distributed according
    to Poisson's law
  \begin{equation}
    \mathbbm{P}\big\{ \omega \in \Omega \, : \, \nu_\omega(\Lambda)= k \big\}
    = \frac{ \big( \overline{\nu}(\Lambda ) \big)^k}{k!} \exp\big[ - \overline{\nu}(\Lambda ) \big], \quad k \in \mathbbm{N}_0
  \end{equation}
  for any bounded $ \Lambda \in \mathcal{B}(\mathbbm{R}^d) $. The case $ \overline{\nu}(\Lambda ) = \varrho | \Lambda | $ corresponds to
  the usual \emph{Poisson process}
  with parameter $ \varrho > 0 $.
 \item[(D)]\label{displace} the \emph{displacement measure} $ \nu = \sum_{j \in \mathbbm{Z}^d} \delta_{j + d_j} $. Here the random variables
   $ d_j \in \Lambda_0 $ are independent and identically distributed
     over the unit cube. The case $ d_j = 0 $ corresponds to the (non-random) \emph{periodic point measure} $ \nu = \sum_{j \in \mathbbm{Z}^d} \delta_{j} $.
\end{indentnummer*}

Any (generalised) Poisson measure~(P) also satisfies Assumption~\ref{prob}. It gives rise to the (generalized) Poissonian
random potential (\ref{eq:defpois}). Unfortunately, Assumption~\ref{prob} is never satisfied for any displacement measure~(D).
However, a corresponding
compound point process $ \nu = \sum_{j \in \mathbbm{Z}^d} q_j \delta_{x_j } $ will satisfy Assumption~\ref{prob} under suitable conditions on the random variables $ (q_j) $.
In order to satisfy Assumption~\ref{intensitymeas}, we take $ (q_j)_{j \in \mathbbm{Z}^d} $ independent and identically distributed, positive random
variables with $  0 < \mathbbm{E}[ q_0 ] < \infty $.

Two examples of such compound point processes, for which  Assumptions~\ref{stat}--\ref{prob} hold, are:
\begin{indentnummer*}
  \item[(P')] the \emph{compound (generalised) Poisson measure} $ \nu = \sum_j q_j \delta_{\xi_{j}} $ with $ (\xi_j) $ as in (P).
  \item[(D')] the \emph{compound displacement measure} $ \nu = \sum_{j \in \mathbbm{Z}^d} q_j \delta_{j + d_j} $ with $ d_j $ as in (D).
    Assumption~\ref{prob} requires
    $ \mathbbm{P}\big\{ \omega \in \Omega : \, q_{\omega,0} \in [0, \varepsilon[ \big\} \geq \varepsilon^\kappa $ for small enough $ \varepsilon > 0 $ and some $ \kappa > 0 $.
    The case $ d_j = 0 $ gives the \emph{alloy-type measure} $  \nu = \sum_{j \in \mathbbm{Z}^d} q_j \delta_{j}  $ associated with the
  alloy-type random potential (\ref{eq:defRP}).
 \end{indentnummer*}

\begin{remark}
We note that in case (P') there are no further requirements on $ (q_j) $.
Moreover, our results in Subsection~\ref{sec:Lif} below also apply to alloy-type random potentials~(\ref{eq:defRP})
with bounded below random variables $ (q_j) $, not only
positive ones. This follows from the fact that one may add $ x \mapsto \sum_{j \in \mathbbm{Z}^d} q_{\rm min} f(x-j) $ to the periodic background
potential $ U_{\rm per} $ (confer (\ref{eq:defH}) and Assumption~\ref{ass:U} below).
\end{remark}

\subsection{Random Sch{\"o}dinger operators and their integrated density of states}
For any of the above defined random potentials $ V $, we study the corresponding random Schr\"o\--dinger operator, which is informally given by the
second order differential
operator
\begin{equation}\label{eq:defH}
  H(V_\omega) :=  - \Delta + U_{\rm per} + V_\omega
\end{equation}
on the Hilbert space $ {\rm L}^2(\mathbbm{R}^d) $ of complex-valued, square-integrable functions on $ \mathbbm{R}^d $.
Thereby the periodic background potential $U_{\rm per} $ (acting in (\ref{eq:defH}) as a multiplication operator) is required to satisfy the following
\begin{assumption}\label{ass:U}
  The background potential $ U_{\rm per} : \mathbbm{R}^d \to \mathbbm{R} $ is $ \mathbbm{Z}^d $-periodic and
$ U_{\rm per} \in {\rm L}_{\rm loc}^p\left(\mathbbm{R}^d\right) $
  for some $ p > d $.
\end{assumption}
Assumptions~\ref{ass:q} and \ref{ass:f} particularly imply \cite[Cor.~V.3.4]{CaLa90} that $ V_\omega \in {\rm
L}^p_{\rm loc}(\mathbbm{R}^d) $ for $ \mathbbm{P} $-almost all $ \omega \in \Omega $ with the same $ p $ as in
Assumption~\ref{pint}. Together with Assumption~\ref{ass:U} this ensures \cite{KirMar83b} that $ H(V_\omega) $ is
essentially self-adjoint on the space $ \mathcal{C}_c^\infty(\mathbbm{R}^d) $ of complex-valued, arbitrarily often
differentiable functions with compact support for $ \mathbbm{P} $-almost all $ \omega \in \Omega $. Since $ V $ is
$ \mathbbm{Z}^d$-ergodic (confer Lemma~\ref{lemma:mix}),
the spectrum of $ H(V_\omega) $ coincides with a non-random set for $ \mathbbm{P}$-almost all $ \omega \in \Omega $  \cite[Thm.~1]{KirMar82}.\\

For any $ d $-dimensional open cuboid $ \Lambda \subset \mathbbm{R}^d $, the restriction of (\ref{eq:defH}) to
$ \mathcal{C}_c^\infty(\Lambda) $ defines a self-adjoint operator $ H_\Lambda^D(V_\omega) $  on $ L^2(\Lambda) $, which corresponds to
taking Dirichlet boundary conditions \cite{ReSi4}. It is bounded below and has purely discrete spectrum with
eigenvalues $\lambda_0(H_\Lambda^D(V_\omega) < \lambda_1(H_\Lambda^D(V_\omega) \leq  \lambda_2(H_\Lambda^D(V_\omega) \leq \ldots$ ordered by magnitude and
repeated according to their multiplicity.
Our main quantity of interest, the integrated density of states, is then defined as the infinite-volume limit
\begin{equation}\label{eq:defN}
   N(E) := \lim_{|\Lambda| \to \infty } \frac{1}{|\Lambda|} \,
   \# \Big\{ n \in \mathbbm{N}_0 \, : \, \lambda_n\left(H_\Lambda^D(V_\omega)\right) < E \Big\}
\end{equation}
More precisely, thanks to the $\mathbbm{Z}^d$-ergodicity of the random potential there is a
set $ \Omega_0 \in \mathcal{A} $ of full probability, $ \mathbbm{P}(\Omega_0) = 1 $, and a non-random unbounded distribution function
$ N : \mathbbm{R} \to [0, \infty [ $
such that (\ref{eq:defN}) holds for all $ \omega \in \Omega_0 $ and all continuity points $ E \in \mathbbm{R} $ of $ N $.
The set of growth points of $ N $ coincides with the almost-sure spectrum of $ H(V_\omega) $, confer \cite{Kir89,CaLa90,PaFi92}. \\

\subsection{Lifshits tails}\label{sec:Lif}
The main result of the present paper generalises the result (\ref{eq:lifexp}) of \cite{KirSim86, Mez87} on the Lifshits exponent
for alloy-type random potentials with isotropically decaying impurity potential $f$
to the case of anisotropic decay and more general random potentials (\ref{eq:defV}).
We note that isotropic decay corresponds to taking $m=1$ in Assumption~\ref{ass:f} or, what is the
same, $\alpha := \alpha_k $ for all $ k \in \{1, \dots , m\} $.
\begin{theorem}\label{Thm:Hauptresultat}
Let $ H(V_\omega) $ be a random Schr\"odinger operator (\ref{eq:defH}) with random potential (\ref{eq:defV}) satisfying Assumptions~\ref{ass:q} and \ref{ass:f}, and
a periodic background potential satisfying Assumption~\ref{ass:U}.
Then its integrated density of states $ N $ drops down to zero exponentially
near $ E_0 := \infspec H(0) $ with Lifshits exponent given by
\begin{equation}\label{eq:Hauptresultat}
  \eta := \lim_{E \downarrow E_0} \frac{\log | \log N(E) | }{| \log (E-E_0) |} = \sum_{k=1}^m \max\left\{ \frac{d_k}{2} , \frac{\gamma_k}{1 - \gamma}\right\},
\end{equation}
where $ \gamma_k := d_k / \alpha_k $ and $ \gamma := \sum_{k=1}^m \gamma_k $.
\end{theorem}
%
\begin{remarks}
\begin{nummer}
  \item
    As a by-product, it turns out that the infimum of the almost-sure spectrum of $ H(V_\omega) $
    coincides with that of $ H(0) = - \Delta + U_{\rm per} $.
      \item
    Thanks to the convention $ 0 = d_k/ \infty \, ( = \gamma_k)  $, Theorem~\ref{Thm:Hauptresultat} remains valid
    if $ \alpha_k = \infty $ for some (or all) $ k \in \{1, \dots , m\} $, confer Remark~\ref{rem:infty}.
  \item
    Assumption~\ref{ass:U} on the local singularities of $ U_{\rm per} $ is slightly more
    restrictive than the one in \cite{KirSim86,Mez87}. It is tailored to ensure certain regularity properties of the ground-state
    eigenfunction of $ H(0) $. As can be inferred from Subsection~\ref{SS:basin} below,
    we may relax Assumption~\ref{ass:U} and require only $ p > d/2 $ (as in \cite{KirSim86,Mez87}) in the interior of the unit cube
    and thus allow for Coulomb singularities there.
  \item
    Even in the isotropic situation $ m = 1 $ Assumption~\ref{ass:f} covers  slightly more impurity potentials
    than in \cite{KirSim86,Mez87}, since we allow
    $ f $ to have zeros at arbitrary large distance from the origin.
  \item
    An inspection of the proof below shows that we prove a slightly better estimate than the double logarithmic asymptotics
    given in (\ref{eq:Hauptresultat}).
    In particular, if the measure $\mu_\omega$ has an atom at zero, more exactly if $\mathbbm{P}\left\{\omega \in  \Omega :  \mu_\omega(\Lambda_0)=0\right\}>0 $,
    then we actually prove
    \begin{equation}
     - C\,\left(E-E_0\right)^\eta \leq \log N(E) \leq- C'\,\left(E-E_0\right)^\eta
    \end{equation}
    for small $E-E_0$.
    This is not quite the logarithmic behaviour (\ref{eq:LifS}) of $N$ since the constants $C>0$ and $C'>0$ do not agree. Note that $\mu_\omega$
    has an atom at zero for the any generalized Poisson measure (P) as well as for a compound displacement measure (D') if
    $\mathbbm{P}\left\{\omega \in  \Omega : q_{\omega,0}(\omega)= 0\right\}>0$.
\end{nummer}
\end{remarks}

For an illustration and interpretation of Theorem~\ref{Thm:Hauptresultat} we consider the special case $ m= 2 $.
The right-hand side of (\ref{eq:Hauptresultat}) then suggests to distinguish the following three cases:\\

\noindent
\begin{tabular}{llcrc}
  {\bf Quantum regime:}  &
    $  \displaystyle \frac{d_1}{2} \geq \frac{\gamma_1}{1 - \gamma} $ & and & $  \displaystyle \frac{d_2}{2} \geq \frac{\gamma_2}{1 - \gamma} $.
    \hfill & {\sf (qm)} \\[2ex]
   {\bf Quantum-classical regime:} &
    $  \displaystyle
       \frac{d_1}{2} \geq \frac{\gamma_1}{1 - \gamma} $ & and & $ \displaystyle \frac{d_2}{2} < \frac{\gamma_2}{1 - \gamma} $   \hfill & {\sf (qm/cl)} \\[2ex]
    \hfill or:   &
            $ \displaystyle
      \frac{d_1}{2} < \frac{\gamma_1}{1 - \gamma} $ & and & $ \displaystyle \frac{d_2}{2} \geq \frac{\gamma_2}{1 - \gamma}$   \hfill & {\sf (cl/qm)} \\[2ex]
  {\bf Classical regime:} &
  $ \displaystyle \frac{d_1}{2} < \frac{\gamma_1}{1 - \gamma} $   & and & $ \displaystyle \frac{d_2}{2} < \frac{\gamma_2}{1 - \gamma} $  \hfill & {\sf (cl)}\\[4ex]
\end{tabular}

In comparison to the result (\ref{eq:lifexp}) for $m=1$
the main finding of this paper is the emergence of a regime corresponding
to mixed quantum and classical character of the Lifshits tail.
A remarkable fact about the Lifshits exponent (\ref{eq:Hauptresultat}) is that
the directions $ k \in \{1, 2\} $ related to the anisotropy do not show up separately as one might expect naively.
In particular, the transition from a quantum to a classical regime for the $ x_k $-direction does not occur if
$ d_k/ 2 = \gamma_k/ (1-\gamma_k) $, but rather if $ d_k/ 2 = \gamma_k/ (1-\gamma) $.
This intriguing intertwining of directions through $ \gamma $ may be interpreted in terms of the marginal impurity potentials
$ f^{(1)} $ and $ f^{(2)} $ defined in (\ref{eq:deff2}) and (\ref{eq:deff1}) below.
In fact, when writing $ \gamma_2/ (1-\gamma) = d_2 /\left( \alpha_2(1-\gamma_1) - d_2\right) $ and identifying $ \alpha_2(1-\gamma_1) $ as the
decay exponent of $ f^{(2)} $ by Lemma~\ref{lemma:margpot1} below, it is clear that $ f^{(2)} $ serves as an effective potential for the
$ x_2 $-direction as far as the quantum-classical transition is concerned.
In analogy, $ f^{(1)} $ serves as the effective potential for the $x_1 $-direction.
Heuristic arguments for the importance of the marginal potentials in the presence of an anisotropy
can be found in \cite{LeWa03}.

\section{Basic inequalities and auxiliary results}
%
%
In order to keep our notation as transparent as possible, we will additionally suppose that
\begin{equation}
  E_0 = 0 \qquad \mbox{and} \qquad m= 2
\end{equation}
throughout the subsequent proof of Theorem~\ref{Thm:Hauptresultat}.
In fact, the first assumption can always be achieved
by adding a constant to $ H(0) $.

The strategy of the proof is roughly the same as in \cite{KirSim86,Mez87}, which in turn is based on \cite{KirMar83a,Sim85}.
We use bounds on the integrated density of states $ N $
and subsequently employ the Rayleigh-Ritz principle and Temple's inequality \cite{ReSi4} to estimate the occurring ground-state energies from above and below.
The basic idea to construct the bounds on $ N $ is to
partition the configuration space $ \mathbbm{R}^d $ into congruent domains and employ some bracketing technique for $ H(V_\omega) $.
The most straightforward of these techniques is Dirichlet or Neumann bracketing.
However, to apply Temple's inequality to the arising Neumann ground-state energy, the authors of \cite{KirSim86} required
that $ U_{\rm per} $ is reflection invariant.
To get rid of this additional assumption, Mezincescu \cite{Mez87} suggested an alternative upper bound on $ N $ which is based on
a bracketing technique corresponding to certain
Robin (mixed) boundary conditions. In his honour, we will refer to these particular Robin boundary conditions as Mezincescu boundary conditions.
\subsection{Mezincescu boundary conditions and basic inequalities}\label{SS:basin}
%
%
Assumption~\ref{ass:U} on $ U_{\rm per} $ implies \cite[Thm.~C.2.4]{Sim82} that there is a continuously differentiable
representative $ \psi : \mathbbm{R}^d \to \, ]0,\infty[ $ of the
strictly positive ground-state eigenfunction of $ H(0) = - \Delta + U_{\rm per} $, which is ${\rm L}^2 $-normalised on the unit cube
$ \Lambda_0 $,
\begin{equation}
  \int_{\Lambda_0}  \psi(x) ^2 dx = 1.
\end{equation}
The function $ \psi $ is $ \mathbbm{Z}^d $-periodic, bounded from below by a strictly positive constant and obeys $ H(0) \psi = E_0 \psi = 0 $.

Subsequently, we denote by $ \Lambda \subset \mathbbm{R}^d $ a $ d $-dimensional, open cuboid which is compatible with the lattice $ \mathbbm{Z}^d $, that is,
we suppose that it coincides with the interior of the union of $ \mathbbm{Z}^d $-translates of the closed unit cube.
On the boundary $ \partial \Lambda $ of $ \Lambda $ we define $ \chi: \partial \Lambda \to \mathbbm{R} $
as the negative of the outer normal derivative of $ \log \psi $,
\begin{equation}\label{eq:defchi}
  \chi(x):= - \frac{1}{\psi(x) }\, \left(n \cdot \nabla\right) \psi (x), \quad x \in \partial \Lambda.
\end{equation}
Since $ \chi \in L^{\infty}(\partial \Lambda) $ is bounded,
the sesquilinear form
\begin{equation}\label{eq:defquad}
  (\varphi_1,\varphi_2) \mapsto \int_\Lambda  \overline{ \nabla \varphi_1(x)} \cdot  \nabla \varphi_2(x) \, dx
  + \int_{\partial \Lambda} \chi(x)  \, \overline{\varphi_1(x)} \varphi_2(x) \, dx,
\end{equation}
with domain $ \varphi_1 $, $ \varphi_2 \in W^{1,2}(\Lambda) := \big\{ \varphi \in L^2(\Lambda) \, : \nabla_j \varphi \in L^2(\Lambda) \; \mbox{for all} \;
j \in \{1, \dots , d\} \big\}$, is symmetric, closed and lower bounded, and thus
uniquely defines a self-adjoint operator $ - \Delta_\Lambda^\chi =: H_\Lambda^\chi(0) - U_{\rm per} $ on
$ L^2(\Lambda) $.
In fact, the condition $ \chi \in L^{\infty}(\partial \Lambda) $ guarantees that boundary term in (\ref{eq:defquad})
is form-bounded with bound zero relative to the first term,
which is just the quadratic form corresponding to the (negative) Neumann Laplacian.
Consequently \cite[Thm. XIII.68]{ReSi4},
both the Robin Laplacian $ - \Delta_\Lambda^\chi $ as well as $ H_\Lambda^\chi(V_\omega) := - \Delta_\Lambda^\chi + U_{\rm per} + V_\omega $, defined as a form sum
on $ W^{1,2}(\Lambda) \subset L^2(\Lambda) $,
have compact resolvents.
Since $ H_\Lambda^\chi(V_\omega) $ generates a positivity preserving semigroup, its ground-state is simple and comes with a
strictly positive eigenfunction \cite[Thm.~XIII.43]{ReSi4}.

\begin{remarks}
  \begin{nummer}
  \item
    In the boundary term in (\ref{eq:defquad}) we took the liberty to denote
    the trace of $ \varphi_j \in W^{1,2}(\Lambda) $ on $ \partial \Lambda $ again by $ \varphi_j $.
  \item
    Partial integration shows that the quadratic form (\ref{eq:defquad}) corresponds to imposing Robin boundary conditions
    $ \left( n \cdot \nabla + \chi \right) \psi |_{\partial \Lambda} = 0 $ on functions $ \psi $ in the domain of the Laplacian on $ \rm {L}^2(\Lambda) $.
    Obviously, Neumann boundary conditions correspond to the special case $ \chi = 0 $. With the present choice (\ref{eq:defchi}) of $ \chi $ they
    arise if $ U_{\rm per} = 0 $ such that $ \psi = 1 $ or,
    more generally, if $ U_{\rm per} $ is reflection invariant (as was supposed in \cite{KirSim86}).
  \item Denoting by
    $ \lambda_0(H_\Lambda^\chi(V_\omega) ) < \lambda_1(H_\Lambda^\chi(V_\omega) ) \leq \lambda_2(H_\Lambda^\chi(V_\omega) ) \leq \dots $
    the eigenvalues of $ H_\Lambda^\chi(V_\omega) $, the eigenvalue-counting function
    \begin{equation}
      N\left(E;  H_\Lambda^\chi(V_\omega)\right)
      := \# \left\{ n \in \mathbbm{N}_0 \, : \, \lambda_n\left(H_\Lambda^\chi(V_\omega)\right) < E \right\}
    \end{equation}
    is well-defined for all $ \omega \in \Omega $ and all energies $ E \in \mathbbm{R} $.
    If $ U_{\rm per} $ is bounded from below, it follows from \cite[Thm.~1.3]{Min02} and (\ref{eq:defN}) that
    $ N(E) = \lim_{|\Lambda| \to \infty }  $\hspace{0pt}$|\Lambda|^{-1} N(E;  H_\Lambda^\chi(V_\omega)) $.
    We also refer to  \cite{Min02} for proofs of some of the above-mentioned properties of the Robin Laplacian.
  \end{nummer}
\end{remarks}
One important point about the Mezincescu boundary conditions (\ref{eq:defchi}) is that the restriction of $ \psi $ to $ \Lambda $
continues to be the ground-state eigenfunction of $ H_\Lambda^\chi(0) $ with eigenvalue $ \lambda_0( H_\Lambda^\chi(0))  =  E_0  =0 $.
This follows from the fact that $ \psi $ satisfies the eigenvalue equation, the boundary conditions and that $ \psi $ is strictly positive.\\

Our proof of Theorem~\ref{Thm:Hauptresultat} is based on the following sandwiching bound on the integrated density of states.

\begin{proposition}\label{prop:basin}
Let $ \Lambda \subset \mathbbm{R}^d $ be a $ d $-dimensional open cuboid, which is compatible with the lattice $ \mathbbm{Z}^{d} $.
Then the integrated density of states $ N $ obeys
\begin{eqnarray}\label{eq:basic}
  && |\Lambda|^{-1}   \, \mathbbm{P}\Big\{ \omega \in \Omega \, : \,  \lambda_0\left( H_\Lambda^D(V_\omega) \right) < E \Big\}
    \leq
  N(E) \nonumber\\
   && \qquad \qquad\quad\leq
  |\Lambda|^{-1} N\left(E; H_\Lambda^\chi(0)\right) \;\; \mathbbm{P}\Big\{  \omega \in \Omega \, : \, \lambda_0\left( H_\Lambda^\chi(V_\omega) \right) < E \Big\}
\end{eqnarray}
for all energies $ E \in \mathbbm{R} $.
\end{proposition}
\begin{proof}
  For the lower bound on $ N $, see \cite[Eq.~(4) and~(21)]{KirMar83a} or \cite[Eq.~(2)]{KirSim86}. The upper bound follows from \cite[Eq.~(29)]{Mez87}.
\end{proof}
\begin{remark}
  Since the bracketing  \cite[Prop.~1]{Mez87} \cite[Probl.~I.7.19]{CaLa90} applies to Robin boundary conditions
  with more general real $ \chi \in {\rm L}^\infty(\partial \Lambda) $  than the one defined
  in (\ref{eq:defchi}), the same is true for the upper bound in (\ref{eq:basic}).
\end{remark}

\subsection{Elementary facts about marginal impurity potentials}
Key quantities in our proof of Theorem~\ref{Thm:Hauptresultat} are the marginal impurity potentials
$ f^{(1)} : \mathbbm{R}^{d_1} \to [0,\infty[ $
and $ f^{(2)} : \mathbbm{R}^{d_2} \to [0,\infty[ $ for the $ x_1 $- and $ x_2 $-direction, respectively. For the given
$ f \in {\rm L}^1(\mathbbm{R}^d) $ they are defined as follows
\begin{eqnarray}
   f^{(1)}(x_1) & := \int_{\mathbbm{R}^{d_2}} f(x_1,x_2) \, dx_2 .\label{eq:deff2} \\
   f^{(2)}(x_2) & := \int_{\mathbbm{R}^{d_1}} f(x_1,x_2) \, dx_1 \label{eq:deff1}
\end{eqnarray}
The aim of this Subsection is to collect properties of $ f^{(2)} $.
Since $ f^{(1)} $ results from $ f^{(2)} $ by exchanging the role of $ x_1 $ and $ x_2 $, analogous properties apply to $ f^{(1)} $.
\begin{lemma}\label{lemma:margpot1}
  Assumption~\ref{ass:f} with $ m = 2 $ implies that there exist two constants $ 0 < f_1 $, $ f_2 < \infty $ such that
  \begin{equation}\label{eq:margpot1}
    \frac{f_1}{|x_2|{}^{\alpha_2 (1- \gamma_1)}} \leq  \int_{|y_2|{} < \frac{1}{2}} \mkern-20mu f^{(2)}(y_2 - x_2)\, dy_2 , \qquad
     f^{(2)}(x_2) \leq \frac{f_2}{|x_2|{}^{\alpha_2 (1- \gamma_1)}}
  \end{equation}
  for large enough $ |x_2|{} > 0 $.
\end{lemma}
\begin{proof}
  The lemma follows by elementary integration. In doing so, one may replace the maximum norm $ | \cdot | $
  by the equivalent Euclidean $ 2 $-norm in both (\ref{eq:assf}) and (\ref{eq:margpot1}).
\end{proof}
\begin{lemma}\label{lemma:margpot2}
 Assumption~\ref{ass:f} with $ m = 2 $ implies that there exists some constant $ 0 < f_3 < \infty $ such that
 \begin{equation}\label{eq:margpot2}
   \int_{|x_2|{} > L } \mkern-20mu f^{(2)}(x_2 ) \, d x_2 \leq f_3 \, L^{-\alpha_2 (1-\gamma)}
 \end{equation}
 for sufficiently large $ L > 0$.
\end{lemma}
\begin{proof}
  By Lemma~\ref{lemma:margpot1} we have
  $  \int_{|x_2|{} > L } f^{(2)}(x_2 ) \, d x_2 \leq f_2 \int_{|x_2|{} > L } |x_2|^{-\alpha_2(1-\gamma_1)} \, d x_2 $
  for sufficiently large $ L > 0 $. The assertion follows
  by elementary integration and the fact that $ \alpha_2 ( 1 - \gamma_1) - d_2 = \alpha_2 (1 - \gamma) $.
\end{proof}
\begin{remark}
One consequence of Lemma~\ref{lemma:margpot2}, which will be useful below, is the following inequality
 \begin{equation}\label{eq:nuetz}
  \sup_{|y_2|{} \leq L/2} \int_{|x_2|{} > L^\beta } \mkern-20mu f^{(2)}(x_2 -y_2) \, d x_2 \leq f_3 \,
   \left(2/L^\beta\right)^{\alpha_2 (1-\gamma)}
 \end{equation}
 valid for all $ \beta \geq 1 $ and sufficiently large $ L > 1$. It is obtained by observing that the integral in (\ref{eq:nuetz}) equals
 \begin{equation}
   \int_{|x_2+j_2|{} > L^\beta } \mkern-20mu f^{(2)}(x_2) \, d x_2 \leq \int_{|x_2|{} \geq L^\beta/2 } \mkern-15mu f^{(2)}(x_2) \, d x_2.
 \end{equation}
 Here the last inequality results from the triangle inequality $ |x_2+y_2|{} \leq |x_2|{} + |y_2|{} $ and
 the fact that $ | y_2 |{}  L/2 \leq L^\beta/2 $.
\end{remark}


\section{Upper bound}
For an asymptotic evaluation of the upper bound in Proposition~\ref{prop:basin} for small energies $ E $
we distinguish the three regimes defined below Theorem~\ref{Thm:Hauptresultat}: quantum,
quantum-classical and classical.
%
\subsection{Regularisation of random Borel measure}
In all of the above mentioned cases it will be necessary to regularise the given random Borel measure $ \mu $ by introducing a cut off.
For this purpose we define
a regularised random Borel measure $ \mu^{(h)} : \Omega \times \mathcal{B}(\mathbbm{R}^d) \to [0, \infty[ $  with parameter $ h > 0 $
by
  $ \mu^{(h)}_{\omega}(\Lambda) := \sum_{j \in \mathbbm{Z}^d} \mu^{(h)}_{\omega}\big(\Lambda \cap \Lambda_j \big) $
  where
  \begin{equation}\label{def:regu}
    \mu^{(h)}_{\omega}\big(\Lambda \cap \Lambda_j \big) :=  \left\{
      \begin{array}{l@{\qquad}l}
        \mu_{\omega}\big(\Lambda \cap \Lambda_j \big) &  \mu_{\omega}\big(\Lambda_j \big) \leq h \\[1ex]
        h \; \displaystyle\frac{\mu_{\omega}\big(\Lambda \cap \Lambda_j \big)}{ \mu_{\omega}\big(\Lambda_j \big)} & \mbox{otherwise}
      \end{array}
      \right.
  \end{equation}
  for all $ \Lambda \in \mathcal{B}(\mathbbm{R}^d) $ and all $ \omega \in \Omega $.
\begin{remark}
Since $ \mu^{(h)}_\omega\left( \emptyset \right) = 0 $
and $  \mu^{(h)}_\omega\big( \bigcup_n \Lambda^{(n)} \big) = \sum_n  \mu^{(h)}_\omega\big( \Lambda^{(n)} \big) $ for any collection of disjoint
$  \Lambda^{(n)}\in \mathcal{B}(\mathbbm{R}^d) $, each realization
$ \mu^{(h)}_\omega $ is indeed a measure on the Borel sets $ \mathcal{B}(\mathbbm{R}^d) $.
It is locally finite and hence a Borel measure, because $ \mu_\omega^{(h)}(\Lambda_j) \leq h $ for all $ j \in \mathbbm{Z}^d $ and all $ \omega \in \Omega $.
\end{remark}
For future reference we collect some properties of $ \mu^{(h)} $.
\begin{lemma}\label{lemma:regu}
  Let $ h > 0 $. Then the following three assertions hold true:
  \begin{indentnummer}
    \item\label{schranke}
      $ \mu^{(h)}_\omega(\Lambda) \leq \min\big\{ \mu_\omega(\Lambda),
      h \; \# \left\{ j \in \mathbbm{Z}^d \, : \, \Lambda \cap \Lambda_j \neq \emptyset \right\}\big\}  \;$
      for all $ \Lambda \in \mathcal{B}(\mathbbm{R}^d) $ and all $ \omega \in \Omega $.
    \item\label{Borel}
      the intensity measure $ \overline{\mu}^{(h)} : \mathcal{B}(\mathbbm{R}^d) \to [0, \infty[ $ given by
          $ \overline{\mu}^{(h)}(\Lambda) := \mathbbm{E}\big[ \mu^{(h)}(\Lambda) \big] $
      is a Borel measure which is $ \mathbbm{Z}^d $-periodic and obeys $ \overline{\mu}^{(h)}(\Lambda_0) > 0 $.
    \item the random variables $ \big( \mu^{(h)}(\Lambda_j) \big)_{j \in \mathbbm{Z}^d} $ are independent and identically distributed.
  \end{indentnummer}
\end{lemma}%
\begin{proof}
The first part of the first assertion is immediate. The other part follows from the monotonicity
$ \mu_\omega(\Lambda \cap \Lambda_j ) \leq  \mu_\omega(\Lambda_j ) \leq h $ for all $ \Lambda \in \mathcal{B}(\mathbbm{R}^d) $,
$ j \in \mathbbm{Z}^d $ and all $ \omega \in \Omega $.
The claimed $ \mathbbm{Z}^d $-periodicity of the intensity measure is traced back to the $ \mathbbm{Z}^d $-stationarity of $ \mu $.
The inequality in the second assertion holds, since $ \mu_\omega(\Lambda_0) $ is not identical zero
for $ \mathbbm{P} $-almost all $ \omega \in \Omega $ (confer Assumption~\ref{ass:q}).
The third assertion follows from the corresponding property of $ \mu $ (confer Assumption~\ref{ass:q}).
\end{proof}

\subsection{Quantum regime}
Throughout this subsection we suppose that {\sf (qm)} holds. Assumption~\ref{ass:f} on the impurity potential requires the existence of
some constant $ f_u > 0 $ and some Borel set $ F  \in \mathcal{B}(\mathbbm{R}^d) $ with $ | F  | > 0 $ such that
\begin{equation}\label{eq:lowerboundqm}
  f \geq f_u \chi_{F}.
\end{equation}
Without loss of generality, we will additionally suppose that $ F \subset \Lambda_0 $.
We start by constructing a lower bound on the lowest Mezincescu eigenvalue $ \lambda_0\left( H_\Lambda^\chi(V_\omega) \right) $
showing up in the right-hand side of (\ref{eq:basic}) when choosing the interior of the closure
\begin{equation}\label{def:cubeqm}
  \Lambda := \overline{\bigcup_{| j |{} < L } \Lambda_{j}}^{\, \rm int}
\end{equation}
of unit cubes, which are at most at a distance $ L > 1 $ from the origin. By construction, the cube $ \Lambda $ is open and compatible with the lattice.
%
%
\subsubsection{Lower bound on the lowest Mezincescu eigenvalue}
From Lemma~\ref{schranke} and (\ref{eq:lowerboundqm}) we conclude that the potential $ V_{\omega, h} : \mathbbm{R}^d \to [0, \infty [ $ given by
\begin{equation}\label{def:Vh}
  V_{\omega, h}(x) := f_u \int_{\mathbbm{R}^d} \chi_{F}(x - y) \, \mu^{(h)}_\omega(dy) = f_u \; \mu^{(h)}_\omega\big(x - F \big)
\end{equation}
in terms of the regularised Borel measure $ \mu_\omega^{(h)} $, provides a lower bound on $ V_\omega $ for every $ h > 0 $ and $ \omega \in \Omega $.
The fact that the pointwise difference $ x - F $ is contained in a cube, which consists of
(at most) $ 3^d $ unit cubes, together with Lemma~\ref{schranke} implies the estimate
\begin{equation}
  V_{\omega, h}(x) \leq 3^d f_u h
\end{equation}
for all $ \omega \in \Omega $ and all $ x \in \mathbbm{R}^d $.
Taking $ h $ small enough thus ensures that
the maximum of the potential $ V_{\omega, h} $ is smaller than the energy difference of the lowest and the first eigenvalue of $  H_{\Lambda}^\chi(0) $.
This enables one to make use of Temple's inequality to obtain a lower bound on the lowest Mezincescu eigenvalue in the quantum regime.
\begin{proposition}\label{prop:Templeqm}
  Let $ \Lambda $ denote the open cube (\ref{def:cubeqm}). Moreover, let
  $ h := \left(r_0 L\right)^{-2} $
  with  $ r_0 > 0 $. Then the lowest eigenvalue of $ H_\Lambda^\chi(V_{\omega, h}^{}) $ is bounded from below
  according to
  \begin{equation}\label{eq:Templeqm}
    \lambda_0\big( H_\Lambda^\chi(V_{\omega, h}^{}) \big)
    \geq \frac{1 }{2 \, | \Lambda | } \int_\Lambda V_{\omega, h}^{}(x) \, \psi(x)^2 \, dx
  \end{equation}
  for all $ \omega \in \Omega $, all $ L > 1 $ and large enough $ r_0 > 0 $.
  {\rm [}Recall the definition of $ \psi $ at the beginning of Subsection~\ref{SS:basin}.{\rm ]}
\end{proposition}
\begin{proof}
  By construction $ \psi_L := |\Lambda|^{-1/2} \, \psi \in L^2(\Lambda) $ is
  the normalised ground-state eigenfunction of $ H_\Lambda^\chi(0) $
  which satisfies $ H_\Lambda^\chi(0) \psi_L = 0 $. Choosing this function as the variational function in
  Temple's inequality \cite[Thm.~XIII.5]{ReSi4} yields the lower bound
  \begin{equation}\label{eq:templeqm}
    \lambda_0\big( H_\Lambda^\chi(V_{\omega, h}^{}) \big) \geq \big\langle \psi_L , V_{\omega, h}^{}  \, \psi_L \big\rangle
    - \frac{\big\langle V_{\omega, h}^{} \,\psi_L , V_{\omega, h}^{}  \, \psi_L \big\rangle}{
      \lambda_1\big( H_\Lambda^\chi(0) \big) - \big\langle \psi_L , V_{\omega, h}^{}  \, \psi_L \big\rangle}
  \end{equation}
  provided the denominator in (\ref{eq:templeqm}) is strictly positive.
  To check this we note that \cite[Prop.~4]{Mez87}
  implies that there is some constant $ c_0 > 0 $ such that
  \begin{equation}
    \lambda_1\big( H_\Lambda^\chi(0) \big) =
    \lambda_1\big( H_\Lambda^\chi(0) \big) - \lambda_0\big( H_\Lambda^\chi(0) \big) \geq 2 c_0 L^{-2}
  \end{equation}
  for all $ L > 1 $.
  Moreover, we estimate
  $  \big\langle \psi_L , V_{\omega, h}^{}  \, \psi_L \big\rangle
    \leq  3^d f_u \, h \leq c_0 L^{-2} $
  for large enough $ r_0 > 0 $.
  To bound the numerator in (\ref{eq:temple}) from above,
  we use the inequality $ \big\langle V_{\omega, h}^{} \,\psi_L , V_{\omega, h}^{}  \, \psi_L \big\rangle
  \leq \big\langle \psi_L , V_{\omega, h}^{}  \, \psi_L \big\rangle \,  3^d f_u h
  \leq  \big\langle \psi_L , V_{\omega, h}^{}  \, \psi_L \big\rangle \, c_0 / (2 L^2) $
  valid for large enough $ r_0 > 0 $.
\end{proof}

We proceed by constructing a lower bound on the right-hand side of (\ref{eq:templeqm}). For this purpose we define the cube
\begin{equation}
  \widetilde \Lambda := \bigcup_{|j |{} < L - 1} \Lambda_j
\end{equation}
which is contained in the cube $ \Lambda $ defined in (\ref{def:cubeqm}). In fact it is one layer of unit cubes smaller than $ \Lambda $.

\begin{lemma}\label{lemma:qm}
  There exists a constant $ 0 < c_1 < \infty $ (which is independent of $ \omega $, $ L $ and $ h $) such that
  \begin{equation}\label{eq:lemmaqm}
    \frac{1}{|\Lambda |} \int_\Lambda V_{\omega, h}^{}(x) \, \psi(x)^2 \, dx
    \geq  \frac{c_1 h}{|\widetilde \Lambda |} \; \# \!\left\{ j \in \mathbbm{Z}^d \cap \widetilde \Lambda \, : \, \mu_\omega\big(\Lambda_j\big) \geq h \right\}
  \end{equation}
  for all $ \omega \in \Omega $, all $ L > 1 $ and all $ h > 0 $.
\end{lemma}
\begin{proof}
Pulling out the strictly positive infimum of $ \psi^2 $ and using its $ \mathbbm{Z}^d $-periodicity, we estimate
  \begin{eqnarray}
  \int_\Lambda V_{\omega, h}^{}(x) \, \psi(x)^2 \, dx & \geq &
 \inf_{z \in \Lambda_0}  \psi(z)^2
 \, f_u \int_{\mathbbm{R}^d} \left| \Lambda \cap (F+y) \right| \, \mu^{(h)}_\omega(dy)\nonumber \\
 & \geq & \inf_{z \in \Lambda_0}  \psi(z)^2 \, f_u \, | F | \; \mu^{(h)}_{\omega}\big(\widetilde \Lambda\big)
\end{eqnarray}
by omitting positive terms and using Fubini's theorem together with the fact that $ F \subset \Lambda_0 $. The proof is completed with the help of the inequality
\begin{equation}
  \mu^{(h)}_{\omega}\big(\widetilde \Lambda\big)  = \sum_{j \in \mathbbm{Z}^d \cap \widetilde \Lambda } \min\big\{ h , \mu_\omega(\Lambda_j) \big\}
  \geq h \; \# \!\left\{ j \in \mathbbm{Z}^d \cap \widetilde \Lambda \, : \, \mu_\omega\big(\Lambda_j\big) \geq h \right\}
\end{equation}
and $ | \Lambda | \leq 3^d | \widetilde \Lambda | $ valid for all $ L > 1 $.
\end{proof}
\subsubsection{Proof of Theorem~\ref{Thm:Hauptresultat} -- first part: quantum regime}
We fix $ r_0 > 0 $ large enough to ensure the validity of (\ref{eq:Templeqm}) in Proposition~\ref{prop:Templeqm}.
For a given energy $ E > 0 $ we then pick
\begin{equation}\label{eq:defLqm1}
 L:= \left(\frac{ c_1 \,}{4 r_0^{2} E}\right)^{1/2}
\end{equation}
where the constant $ c_1 $ has been fixed in Lemma~\ref{lemma:qm}.
Finally, we choose the cube
$ \Lambda $ from (\ref{def:cubeqm}) and set $ h := (r_0 L)^{-2} $.
Proposition~\ref{prop:Templeqm} and (\ref{eq:lemmaqm}) yield the estimate
\begin{eqnarray}
  && \mathbbm{P}\Big\{ \omega \in \Omega \, : \, \lambda_0\left( H_\Lambda^\chi(V_\omega) \right) < E \Big\} \nonumber \\
  && \qquad
  \leq  \mathbbm{P}\Bigg\{ \omega \in \Omega \, : \, \# \!\left\{ j \in \mathbbm{Z}^d \cap \widetilde \Lambda \, : \, \mu_\omega\big(\Lambda_j\big) \geq h \right\}
  < \frac{2 E}{c_1 h} |\widetilde \Lambda | \Bigg\} \nonumber \\
  && \qquad  = \mathbbm{P}\Bigg\{ \omega \in \Omega \, : \, \# \!\left\{ j \in \mathbbm{Z}^d \cap \widetilde \Lambda \, : \, \mu_\omega\big(\Lambda_j\big) < h \right\}
  > \frac{| \widetilde \Lambda |}{2} \Bigg\}.\quad \label{eq:largedevqm}
\end{eqnarray}
Here the last equality uses the fact that $ h = 4 E /c_1 $. In case $ \overline{\mu}(\Lambda_j) > h $, that is, for sufficiently small $ E $, the right-hand side
is the probability of a large deviation event \cite{DeZe98}.
Consequently (confer~\cite[Prop.~4]{KirSim86}), there exists a constant $ 0 < c_2 < \infty $, such that (\ref{eq:largedevqm}) is estimated
from above by
\begin{equation}
  \exp\left[- c_2 | \widetilde \Lambda |\right] \leq \exp\left[ - c_2 n_u L^{d} \right] = \exp\left[ - c_3 E^{-d/2} \right]
\end{equation}
Here the inequality follows from the estimate $ | \widetilde \Lambda | \geq n_u L^d $ for some constant $ n_u > 0 $ and all $ L > 2 $. The
existence of a constant $ c_3 > 0 $ ensuring the validity of the last equality follows from (\ref{eq:defLqm1}).
Inserting this estimate in the right-hand side of (\ref{eq:basic})
completes the first part of the proof of Theorem~\ref{Thm:Hauptresultat} for the quantum-classical regime, since the pre-factor in the upper bound in
Proposition~\ref{prop:basin} is negligible. \qed

%
\subsection{Quantum-classical regime}
Without loss of generality we suppose that {\sf (qm/cl)} holds throughout this subsection, that is
 $ d_1/2 \geq \gamma_1/(1 - \gamma) $ and $ d_2/ 2 < \gamma_2/(1 - \gamma) $.
We start by constructing a lower bound on the lowest Mezincescu eigenvalue $ \lambda_0\left( H_\Lambda^\chi(V_\omega) \right) $
showing up in the right-hand side of (\ref{eq:basic}) when choosing
\begin{equation}\label{def:cuboid}
  \Lambda := \overline{\bigcup_{| j_1 |{} < L } \Lambda_{(j_1,0)}}^{\, \rm int}
\end{equation}
a cuboid with some $ L > 1 $. By construction it is open and compatible with the lattice.
%
%
\subsubsection{Lower bound on the lowest Mezincescu eigenvalue}
%
%
From Lemma~\ref{schranke} we conclude that for every $ R > 0 $ and $ \omega \in \Omega $ the potential
$V_{\omega, R}^{}: \mathbbm{R}^d \to [0, \infty [ $ given by
\begin{equation}\label{def:VL}
  V_{\omega, R}^{}(x) :=\int_{| y_2 |{} > R }\mkern-20mu  f(x-y) \, \mu^{(1)}_{\omega}(dy)
\end{equation}
in terms of the regularised Borel measure $ \mu_\omega^{(1)} $, provides a lower bound on $ V_\omega $.
Therefore
$ \lambda_0\left( H_\Lambda^\chi(V_\omega) \right) \geq \lambda_0\big( H_\Lambda^\chi(V_{\omega, R}^{}) \big) $.
It will be useful to collect some facts related to $ V_{\omega, R} $.
\begin{lemma}\label{lemma:VR}
  Let $ R > 1 $ and define $ V_R^{} : \mathbbm{R}^d \to [0, \infty [ $
  by
  \begin{equation}\label{eq:defV1}
    V_R^{} (x) := \sum_{\substack{j_1 \in \mathbbm{Z}^{d_1} \\ | j_2 |{} > R -1}}  \sup_{y \in \Lambda_j} f(x -y).
  \end{equation}
  Then the following three assertions hold true:
  \begin{indentnummer}
    \item
      $ V_{\omega, R}^{} \leq V_R^{} $  for every $ \omega \in \Omega $.
    \item
      $ V_R^{} $ is $ \mathbbm{Z}^{d_1}$-periodic with respect to translations in the $ x_1 $-direction.
    \item
      there exists some constant $ c > 0 $ such that
      $  \sup_{ x \in \Lambda_0}  V_R^{}(x) \leq c\, R^{-\alpha_2 (1-\gamma)} $
      for large enough $ R > 1 $.
  \end{indentnummer}
\end{lemma}
\begin{proof}
The first assertion follows from the inequalities
\begin{equation}
  V_{\omega, R}^{}(x) \leq \sum_{\substack{j_1 \in \mathbbm{Z}^{d_1} \\ | j_2 |{} > R -1}} \int_{ \Lambda_j} f(x-y) \,  \mu^{(1)}_{\omega}(dy)
\end{equation}
and $ \mu^{(1)}_{\omega}(\Lambda_j) \leq 1 $ valid for all $ \omega \in  \Omega $.
The second assertion holds true by definition.
The third assertion derives from (\ref{eq:assf}) and is the ``summation'' analogue of Lemma~\ref{lemma:margpot2}.
\end{proof}
The cut-off $ R $ guarantees that the potential $ V_{\omega, R}^{} $
does not exceed a certain value. In particular, taking $ R $ large enough ensures that
this value is smaller than the energy difference of the lowest and the first eigenvalue of $  H_{\Lambda}^\chi(0) $.
This enables one to make use of Temple's inequality to obtain a lower bound on the lowest Mezincescu eigenvalue in the quantum-classical regime.
\begin{proposition}\label{prop:Temple}
  Let $ \Lambda $ denote the cuboid (\ref{def:cuboid}). Moreover, let
  $  R := \left(r_0 L\right)^{2/\alpha_2 (1-\gamma)} $
  with  $ r_0 > 0 $. Then the lowest eigenvalue of $ H_\Lambda^\chi(V_{\omega, R}^{}) $ is bounded from below
  according to
  \begin{equation}\label{eq:Temple}
    \lambda_0\big( H_\Lambda^\chi(V_{\omega, R}^{}) \big)
    \geq \frac{1 }{2 \, | \Lambda | } \int_\Lambda V_{\omega, R}^{}(x) \, \psi(x)^2 \, dx
  \end{equation}
  for all $ \omega \in \Omega $, all $ L > 1 $ and large enough $ r_0 > 0 $.
  {\rm [}Recall the definition of $ \psi $ at the beginning of Subsection~\ref{SS:basin}.{\rm ]}
\end{proposition}
\begin{proof}
  The proof parallels the one of Proposition~\ref{prop:Temple}.
  By construction $ \psi_L := |\Lambda|^{-1/2} \, \psi \in L^2(\Lambda) $ is
  the normalised ground-state eigenfunction of $ H_\Lambda^\chi(0) $
  which satisfies $ H_\Lambda^\chi(0) \psi_L = 0 $. Choosing this function as the variational function in
  Temple's inequality \cite[Thm.~XIII.5]{ReSi4} yields the lower bound
  \begin{equation}\label{eq:temple}
    \lambda_0\big( H_\Lambda^\chi(V_{\omega, R}^{}) \big) \geq \big\langle \psi_L , V_{\omega, R}^{}  \, \psi_L \big\rangle
    - \frac{\big\langle V_{\omega, R}^{} \,\psi_L , V_{\omega, R}^{}  \, \psi_L \big\rangle}{
      \lambda_1\big( H_\Lambda^\chi(0) \big) - \big\langle \psi_L , V_{\omega, R}^{}  \, \psi_L \big\rangle}
  \end{equation}
  provided the denominator in (\ref{eq:temple}) is strictly positive.
  To check this we note that a simple extension of \cite[Prop.~4]{Mez87} from cubes to cuboids
  implies that there is some constant $ c_0 > 0 $ such that
  $  \lambda_1\big( H_\Lambda^\chi(0) \big) =
    \lambda_1\big( H_\Lambda^\chi(0) \big) - \lambda_0\big( H_\Lambda^\chi(0) \big) \geq 2 c_0 L^{-2}
  $  for all $ L > 1 $.
  Moreover, using Lemma~\ref{lemma:VR} and the definition of $ R $ we estimate
  \begin{equation}
    \big\langle \psi_L , V_{\omega, R}^{}  \, \psi_L \big\rangle
    \leq  \big\langle \psi_L , V_{R}^{}  \, \psi_L \big\rangle
    =    \int_{\Lambda_0}  V_{R}^{}(x) \,\psi(x)^2 dx  \leq c \, \big( r_0 L\big)^{-2} \leq c_0 L^{-2}
  \end{equation}
  for large enough $ r_0 > 0 $.
  To bound the numerator in (\ref{eq:temple}) from above,
  we use the inequality $ \big\langle V_{\omega, R}^{} \,\psi_L , V_{\omega, R}^{}  \, \psi_L \big\rangle
  \leq \big\langle \psi_L , V_{\omega, R}^{}  \, \psi_L \big\rangle \, \sup_{x \in \Lambda} V_R^{}(x) $.
  Lemma~\ref{lemma:VR} ensures that
  $ \sup_{x \in \Lambda} V_R^{}(x) = \sup_{x \in\Lambda_0} V_R^{}(x) $ and thus yields
  the bound
  \begin{equation}
    \big\langle V_{\omega, R}^{} \,\psi_L , V_{\omega, R}^{}  \, \psi_L \big\rangle
     \leq \big\langle \psi_L , V_{\omega, R}^{}  \, \psi_L \big\rangle \,  c \, \big(r_0 L \big)^{-2}
     \leq \big\langle \psi_L , V_{\omega, R}^{}  \, \psi_L \big\rangle \, \frac{c_0}{2} L^{-2}
  \end{equation}
  for large enough $ r_0 > 0 $.
\end{proof}

We proceed by constructing a lower bound on the right-hand side of (\ref{eq:Temple}). For this purpose we set
\begin{equation}
    \widetilde \Lambda := \bigcup_{\substack{ |j_1|{} \leq L/8 \\ R < | j_2 |{} \leq 2 R }} \Lambda_j
  \end{equation}
a union of disjoint cuboids.
%
\begin{lemma}\label{lemma:weiter}
  There exist two constants $ 0 < c_2 $, $ c_3 < \infty $ (which are independent of $ \omega $, $ L $ and $ R $) such that
  \begin{equation}\label{eq:weiter}
     \int_\Lambda V_{\omega, R}^{}(x) \, \psi(x)^2 \, dx \geq  \frac{c_2}{ R^{\alpha_2 (1 - \gamma_1)}}\;
     \mu^{(1)}_{\omega}\big(\widetilde \Lambda\big)
       - c_3 \, | \Lambda | \, L^{-\alpha_1 (1 - \gamma)}
  \end{equation}
  for all $ \omega \in \Omega $ and large enough $ L > 1 $ and $ R > 1 $.
\end{lemma}
\begin{remark}\label{rem:weiter}
An important consequence of this lemma reads as follows.
There exists some constant $ n_u > 0 $ such that the number of lattice points in $ \widetilde \Lambda $
is estimated from below by $ |\widetilde \Lambda | \geq n_u \, |\Lambda| R^{d_2} $
for all $ L > 1 $ and $ R > 1 $ and some constant $ n_u > 0 $.
Therefore
$ |\widetilde \Lambda | / (|\Lambda| R^{\alpha_2 (1 - \gamma_1)}) \geq n_u / R^{\alpha_2 (1 - \gamma)} $.
Choosing $ R = (r_0 L)^{2/\alpha_2 (1-\gamma)} $ as in
Proposition \ref{prop:Temple}, we thus arrive at the lower bound
\begin{equation}\label{eq:weiter2}
     \frac{1}{|\Lambda |} \int_\Lambda V_{\omega, R}^{}(x) \, \psi(x)^2 \, dx \geq
     \frac{c_2 \, n_u }{(r_0 L)^2} \; \frac{1}{|\widetilde \Lambda |}
     \sum_{j \in \mathbbm{Z}^d \cap \widetilde \Lambda} \mu^{(1)}_{\omega}\big(\Lambda_j\big)
       - c_3 \, L^{-\alpha_1 (1 - \gamma)}
\end{equation}
valid for all $ r_0 > 0 $ and large enough $ L > 1 $.
\end{remark}
\begin{proof}[Proof of Lemma~\ref{lemma:weiter}]
  Pulling out the strictly positive infimum of $ \psi^2 $ and using its $ \mathbbm{Z}^d $-periodicity, we estimate
  \begin{eqnarray}
  \int_\Lambda V_{\omega, R}^{}(x) \, \psi(x)^2 \, dx & \geq &
 \inf_{z \in \Lambda_0}  \psi(z)^2
 \int_\Lambda V_{\omega, R}^{}(x) \, dx \nonumber \\
 & \geq & \inf_{z \in \Lambda_0}  \psi(z)^2 \int_{\widetilde \Lambda} \left( \int_\Lambda f(x-y) \, dx \right) \mu^{(1)}_{\omega}(dy)
\end{eqnarray}
by omitting positive terms and using Fubini's theorem.
The inner integral in the last line
is estimated from below with the help of Lemma~\ref{lemma:margpot1} in terms of the marginal impurity potential $ f^{(2)} $
(recall definition (\ref{eq:deff1})) according to
\begin{eqnarray}
 \int_\Lambda f(x-y) \, dx & = & \int_{| x_2 |{} < \frac{1}{2}} \mkern-20mu f^{(2)}(x_2-y_2) \, dx_2
 - \sum_{|k_1|{} \geq L} \int_{\Lambda_{(k_1,0)}}  \mkern-20mu f(x-y) \, dx \nonumber\\
 & \geq & \frac{f_1}{(2 R+1)^{\alpha_2(1-\gamma_1)}} - \sum_{|k_1|{} \geq L} \int_{\Lambda_0}  \! f\big(x+(k_1,0)-y\big) \, dx \label{eq:second}
\end{eqnarray}
for all $ | y_2 |{} \leq 2 R +1 $ and large enough $ R > 0 $.
The first term on the right-hand side yields the first term on the right-hand side of (\ref{eq:weiter}).
To estimate the remainder we decompose the $y$-integration of the second term in (\ref{eq:second}) with respect to $ \mu^{(1)}_{\omega} $ and use the fact that
$ \mu^{(1)}_{\omega}(\Lambda_j) \leq 1 $. This yields an estimate of the form
\begin{eqnarray}
\int_{\widetilde \Lambda}  \left(\int_{\Lambda_0}  \! g(x-y) \, dx \right) \mu^{(1)}_{\omega}(dy)
& \leq & \sum_{j \in \mathbbm{Z}^d \cap \widetilde \Lambda} \; \sup_{y \in \Lambda_0}
\int_{\Lambda_0}  \! g(x-y-j) \, dx \nonumber \\
& \leq & 3^d \sum_{\substack{ |j_1|{} \leq L/2 \\ j_2 \in \mathbbm{Z}^{d_2}}} \int_{\Lambda_0}  \! g(x-j) \, dx \nonumber \\
& = & 3^d \sum_{|j_1|{} \leq L/4} \int_{| x_1 |{} < 1/2} \mkern-10mu g^{(1)}(x_1 - j_1) \, dx_1
\end{eqnarray}
valid for all $ g \in {\rm L}^1(\mathbbm{R}^d) $. Here the second inequality holds for every $ L \geq 8 $ (so that $ L/4 - L/8 \geq 1 $) and
follows from enlargening the $ j_2$-summation and the fact that the pointwise
difference $ \Lambda_0 - \Lambda_0 $ is contained in the cube centred at the origin and consisting of $3^d $ unit cubes.
The last equality uses the definition (\ref{eq:deff2}) for a marginal impurity potential. Substituting $ g (x ) = f (x + (k_1,0)) $
in the above chain of inequalities, performing the $ k_1 $-summation and enlargening the $ x_1 $-integration thus yields
\begin{equation}
3^d \sum_{|j_1|{} \leq L/4 }  \int_{| x_1 |{} > L/2} \mkern-15mu f^{(1)}(x_1 - j_1) \, dx_1
\leq 3^d n_0 | \Lambda | \, \sup_{|j_1 |{} \leq L/4} \int_{| x_1 |{} > L/2} \mkern-15mu f^{(1)}(x_1 - j_1) \, dx_1
\end{equation}
as an upper bound for the remainder for all $ L \geq 8 $.
Here the inequality follows from the estimate $ \#\{ |j_1 |{} \leq L/2 \} \leq n_0 | \Lambda | $ for some $ n_0 < \infty $ and all
$ L > 1$. The proof is completed by employing a result for $ f^{(1)} $ analogous to (\ref{eq:nuetz}).
\end{proof}
%
\subsubsection{Proof of Theorem~\ref{Thm:Hauptresultat} -- first part: quantum-classical regime}
We fix $ r_0 > 1/( 2 \, \overline{\mu}^{(1)}(\Lambda_0) )$ large enough to ensure the validity of (\ref{eq:Temple}) in Proposition~\ref{prop:Temple}.
For a given energy $ E > 0 $ we then pick
\begin{equation}\label{eq:defLqm}
 L:= \left(\frac{ c_2 \, n_u }{2 r_0^{3} E}\right)^{1/2}
\end{equation}
where the constants $ c_2 $ and $ n_u $ have been fixed in Lemma~\ref{lemma:weiter} and Remark~\ref{rem:weiter}.
Finally, we choose the cuboid
$ \Lambda $ from (\ref{def:cuboid}) and set $ R := (r_0 L)^{2/\alpha_2 (1-\gamma)} $.
Proposition~\ref{prop:Temple} and (\ref{eq:weiter2}) then yield the estimate
\begin{eqnarray}
  && \mathbbm{P}\Big\{ \omega \in \Omega \, : \, \lambda_0\left( H_\Lambda^\chi(V_\omega) \right) < E \Big\} \nonumber  \\
  \quad && \leq
  \mathbbm{P}\Bigg\{ \omega \in \Omega \, : \, \frac{1}{|\widetilde \Lambda |}
     \sum_{j \in \mathbbm{Z}^d \cap \widetilde \Lambda} \mu^{(1)}_{\omega}\big(\Lambda_j\big) <
    \frac{(r_0 L)^2}{c_2 n_u} \left( 2 E + c_2 L^{-\alpha_1 (1 - \gamma)} \right) \Bigg\} \nonumber \\
   \quad &&  \leq
  \mathbbm{P}\Bigg\{ \omega \in \Omega \, : \, \frac{1}{|\widetilde \Lambda |}
     \sum_{j \in \mathbbm{Z}^d \cap \widetilde \Lambda} \mu^{(1)}_{\omega}\big(\Lambda_j\big) <
    \frac{2}{r_0} \Bigg\} \label{eq:largedev}
\end{eqnarray}
provided $ E > 0 $ is small enough, equivalently $ L $ is large enough.
Here the last inequality results from (\ref{eq:defLqm}) and
from the first inequality in $\mbox{\sf (qm/cl)}$, which implies that
$ c_3 r_0^3 L^2 \leq c_2 n_u L^{\alpha_1 (1 - \gamma)} $ for large enough $ L > 0 $.
Since $ 2/ r_0 \leq \overline{\mu}^{(1)}(\Lambda_0) $
by assumption on $ r_0$, the right-hand side of (\ref{eq:largedev}) is the probability of a large-deviation event \cite{Dur96,DeZe98}.
Consequently, there exists some constant $ c_4 > 0 $ (which is independent
of $ L $) such that (\ref{eq:largedev}) is estimated from above by
\begin{eqnarray}
  \exp\Big[ - c_4 \, |\widetilde \Lambda | \Big]
  & \leq &\exp\left[ - c_4 \, n_u \, L^{d_1} \left(r_0 L \right)^{2 \gamma_2/(1-\gamma)} \right] \nonumber \\
  & = & \exp\left[ - c_5 \, E^{-d_1/2 - \gamma_2/(1-\gamma)} \right].
\end{eqnarray}
Here the existence of a constant $ c_5 > 0 $ ensuring the validity of the last equality follows from (\ref{eq:defLqm}).
Inserting this estimate in the right-hand side of (\ref{eq:basic})
completes the first part of the proof of Theorem~\ref{Thm:Hauptresultat} for the quantum-classical regime, since the pre-factor in the upper bound in
Proposition~\ref{prop:basin} is negligible. \qed

%
\subsection{Classical regime}
Throughout this Subsection we suppose that $ \mbox{\sf (cl)} $ holds. For an asymptotic evaluation of the upper bound
in Proposition~\ref{prop:basin} in the present case, we define
\begin{equation}\label{eq:defbetacl}
  \beta_k := \frac{2}{d_k}\, \frac{\gamma_k}{1-\gamma} = \frac{2}{\alpha_k \,(1-\gamma)}, \qquad k \in \{1,2\}
\end{equation}
and construct a lower bound on the lowest Mezincescu eigenvalue $ \lambda_0\big( H_{\Lambda_0^{\rm int}}^\chi(V_\omega) \big) $
showing up in the right-hand side of (\ref{eq:basic}) when choosing $ \Lambda = \Lambda_0^{ \rm int} $ the open unit cube there.
%
\subsubsection{Lower bound on the lowest Mezincescu eigenvalue}
For every $ L > 1 $ and $ \omega \in \Omega $
the potential $V_{\omega, L} : \mathbbm{R}^d \to [0, \infty[ $ given by
\begin{equation}\label{def:V}
  V_{\omega, L}(x) := \int_{\substack{|y_1 |{} > L^{\beta_1} \\ | y_2 |{} > L^{\beta_2}}} \! f(x-y) \, \mu^{(1)}_{\omega}(dy)
\end{equation}
in terms of the regularised Borel measure $ \mu_\omega^{(1)} $, provides a lower bound on $ V_\omega $.
Therefore $ \lambda_0\big( H_{\Lambda_0^{\rm int}}^\chi(V_\omega) \big) \geq \lambda_0\big( H_{\Lambda_0^{\rm int}}^\chi(V_{\omega, L}) \big) $.
It will be useful to collect some facts related to $ V_{\omega, L} $.
\begin{lemma}\label{lemma:VL}
 Let $ L > 1 $ and define $ V_L : \mathbbm{R}^d \to [0, \infty [ $
  by
  \begin{equation}
    V_L(x) :=  \sum_{\substack{|j_1 |{} > L^{\beta_1}-1 \\ | j_2 |{} > L^{\beta_2}-1}} \sup_{y \in \Lambda_j} f(x -y).
  \end{equation}
  Then we have $ V_{\omega, L} \leq V_L $ for every $ \omega \in \Omega $. Moreover,
  the supremum
  $ \sup_{x \in \Lambda_0}  V_L(x) $ is arbitrarily small
  for large enough $ L > 1 $.
\end{lemma}
\begin{proof}
  The first assertion follows analogously as in Lemma~\ref{lemma:VR}. The second one derives from the second inequality in (\ref{eq:assf}).
\end{proof}
\begin{remark}
  It is actually not difficult to prove that there exists some constant $ 0< C < \infty $ (which is independent of $ L $) such that
  $ \sup_{x \in \Lambda_0}  V_L(x) \leq C \, L^{-2} $ for large enough $ L > 0 $.
\end{remark}

The next proposition contains the key estimate on the lowest Mezincescu eigenvalue in the classical regime.
In contrast to the quantum-classical regime, the specific choice of the cut-off made in (\ref{def:V}) is irrelevant
as far as the applicability of Temple's inequality in the subsequent Proposition is concerned.
The chosen length scales $ L^{\beta_1} $ and $ L^{\beta_2} $ will rather become important later on.
\begin{proposition}\label{prop:Templecl}
  Let $ \Lambda_0^{\rm int} $ be the open unit cube.
  Then the lowest eigenvalue of $ H_{\Lambda_0^{\rm int}}^\chi(V_{\omega, L}) $ is bounded from below
  according to
  \begin{equation}\label{eq:T}
    \lambda_0\big( H_{\Lambda_0^{\rm int}}^\chi(V_{\omega, L}) \big)
    \geq \frac{1}{2} \int_{\Lambda_0} V_{\omega, L}(x) \, \psi(x)^2 \, dx
  \end{equation}
  for all $ \omega \in \Omega $ and large enough $ L > 1 $.
  {\rm [}Recall the definition of $ \psi $ at the beginning of Subsection~\ref{SS:basin}.{\rm ]}
\end{proposition}
\begin{proof}
  The proof again parallels that of Proposition~\ref{prop:Templeqm}.
  In a slight abuse of notation, let $ \psi $ denote the restriction of $ \psi $ to $ \Lambda_0^{\rm int} $ throughout this proof.
  Temple's inequality \cite[Thm.~XIII.5]{ReSi4} together with the fact that $ H_{\Lambda_0^{\rm int}}^\chi(0) \psi = 0 $ yields the lower bound
  \begin{equation}\label{eq:templecl}
    \lambda_0\big( H_{\Lambda_0^{\rm int}}^\chi(V_{\omega, L}) \big) \geq \left\langle \psi , V_{\omega, L}  \, \psi \right\rangle
    - \frac{\left\langle V_{\omega, L} \,\psi , V_{\omega, L}  \, \psi \right\rangle}{
      \lambda_1\big( H_{\Lambda_0^{\rm int}}^\chi(0) \big) - \left\langle \psi , V_{\omega, L}  \, \psi \right\rangle}
  \end{equation}
  provided that the denominator is strictly positive. To check this we
  employ Lemma~\ref{lemma:VL} and take $ L > 1 $ large enough such that
  $  \left\langle \psi , V_{\omega, L}  \, \psi\right\rangle
      \leq \lambda_1\big( H_{\Lambda_0^{\rm int}}^\chi(0) \big)/ 2 $.
  (Note that $ \lambda_1\big( H_{\Lambda_0^{\rm int}}^\chi(0) \big) $ is independent of $ L $.)
  To estimate the numerator in (\ref{eq:templecl}) from above,
  we use the bound $ \left\langle V_{\omega, L} \,\psi, V_{\omega, L}  \, \psi \right\rangle
  \leq \left\langle \psi , V_{\omega, L}  \, \psi \right\rangle \, \sup_{x \in \Lambda_0} V_L(x) $. Together with
  Lemma~\ref{lemma:VL} this yields
  $  \left\langle V_{\omega, L} \,\psi , V_{\omega, L}  \, \psi \right\rangle
     \leq \left\langle \psi , V_{\omega, L}  \, \psi \right\rangle $\hspace{0pt}$ \lambda_1\big( H_{\Lambda_0^{\rm int}}^\chi(0) \big)/ 4 $
  for large enough $ L > 1 $.
\end{proof}
\begin{remark}
  The simple lower bound $ \lambda_0\big( H_{\Lambda_0^{\rm int}}^\chi(V_{\omega,L}) \big) \geq \inf_{x \in \Lambda_0} V_{\omega,L}(x) $, which was employed
  in \cite{KirSim86}, would yield a result similar to (\ref{eq:lde}) below, but at the price of assuming that the lower bound in (\ref{eq:assf}) holds pointwise.
\end{remark}
We proceed by constructing a lower bound on the right-hand side of (\ref{eq:T}).
For this purpose we set
\begin{equation}
  \widetilde \Lambda :=
  \bigcup_{\substack{2L^{\beta_1} < |j_1 |{} \leq 4 L^{\beta_1}\\ 2L^{\beta_2} < | j_2 |{} \leq 4 L^{\beta_2}}}
  \Lambda_j
\end{equation}
an annulus-shaped region.
\begin{lemma}\label{lemma:runter2}
There exists a constant $ c_6 > 0 $ (which is independent of $ \omega $ and $ L $) such that
\begin{equation}\label{eq:runter}
  \int_{\Lambda_0} V_{\omega, L}(x) \, \psi(x)^2 \, dx  \geq \frac{c_6}{L^{2/(1-\gamma)}} \;
  \mu^{(1)}_{\omega}\big(\widetilde \Lambda\big)
\end{equation}
for large enough $ L > 0 $.
\end{lemma}
\begin{proof}
  Pulling out the strictly positive infimum of $ \psi^2 $, using Fubini's theorem and omitting a positive term, we estimate
  \begin{equation}
    \int_{\Lambda_0} V_{\omega, L}(x) \, \psi(x)^2 \, dx  \geq
     \inf_{z\in \Lambda_0} \psi(z)^2
     \int_{\widetilde \Lambda} \left(\int_{\Lambda_0} f(x-y) \, dx \right) \mu^{(1)}_{\omega}(dy).
  \end{equation}
  Assumption~\ref{ass:f} implies that the estimate
  $ \int_{\Lambda_0} f(x-y) \, dx \geq f_u/ \big[ (3L^{\beta_1})^{\alpha_1} + (3 L^{\beta_2})^{\alpha_2}\big] $
  holds for all $ y \in \widetilde \Lambda $ and large enough $ L > 1$.
  This completes the proof, since
  $ \alpha_k \beta_k = 2/(1-\gamma) $ for both $ k \in \{1,2\} $.
\end{proof}
\begin{remark}\label{rem:runter2}
  There exists some constant $ n_u > 0 $ such that the number of lattice points in $ \widetilde \Lambda $
  can be bounded from below according to
  $ | \widetilde \Lambda | \geq n_u L^{\beta_1 d_1 + \beta_2 d_2} = n_u L^{2\gamma/(1-\gamma)} $ for all $ L > 1 $.
  Lemma~\ref{lemma:runter2} thus implies the inequality
  \begin{equation}\label{eq:runter2}
    \int_{\Lambda_0} V_{\omega, L}(x) \, \psi(x)^2 \, dx  \geq \frac{c_6\, n_u}{L^2} \; | \widetilde \Lambda |^{-1} \,
     \mu^{(1)}_{\omega}\big(\widetilde \Lambda\big)
  \end{equation}
  for large enough $ L > 1$.
\end{remark}
\subsubsection{Proof of Theorem~\ref{Thm:Hauptresultat} -- first part: classical regime}
For a given energy $ E > 0 $ we let
$ L := \left(c_6 n_u \, \overline{\mu}^{(1)}(\Lambda_0) / 4 E\right)^{1/2} $, where the constant $ c_6 $ and $ n_u $ have been
fixed in Lemma~\ref{lemma:runter2} and Remark~\ref{rem:runter2}.
Proposition~\ref{prop:Templecl} and Equation~(\ref{eq:runter2}) then yield the estimate
\begin{equation}
   \mathbbm{P}\Big\{ \omega \in \Omega \, : \, \lambda_0\left( H_{\Lambda_0^{\rm int}}^\chi(V_\omega) \right) < E \Big\}
    \leq
  \mathbbm{P}\Bigg\{ \omega \in \Omega \, : \, \frac{1}{| \widetilde \Lambda |} \,
  \sum_{j \in \mathbbm{Z}^d \cap \widetilde \Lambda} \mu^{(1)}_{\omega}\big( \Lambda_j \big) <
    \frac{2 E \, L^2}{c_6 \, n_u}  \Bigg\} \label{eq:lde}
\end{equation}
provided $ E > 0 $ is small enough, equivalently $ L $ is large enough.
Since $ 2 E  L^2/ $\hspace{0pt}$c_6  n_u = \mathbbm{E}\big[\mu^{(1)}_{\omega}( \Lambda_0)\big]/2 $ and the random variables are
independent and identically distributed,
the last probability is that of a large deviation event \cite{Dur96,DeZe98}.
Consequently, there exists
some $ c_7 > 0 $ such that the right-hand side of (\ref{eq:lde}) is bounded from above by
\begin{eqnarray}\label{eq:finshcl}
  \exp\Big[ - c_7 \, | \widetilde \Lambda | \Big] & \leq &\exp\Big[ - c_7 n_u L^{2\gamma/(1-\gamma)} \Big] \nonumber \\
                             & = &\exp\Big[ - c_7 n_u
                             \left(c_6 n_u\, \overline{\mu}^{(1)}(\Lambda_0)/ 4 E\right)^{\gamma/(1-\gamma)} \Big].
\end{eqnarray}
Since the pre-factor in the upper bound in
Proposition~\ref{prop:basin} is negligible, inserting (\ref{eq:lde}) together with (\ref{eq:finshcl})
in the right-hand side of (\ref{eq:basic})
completes the first part of the proof of Theorem~\ref{Thm:Hauptresultat} for the classical regime. \qed
%
%

\section{Lower bound}
%
To complete the proof of Theorem~\ref{Thm:Hauptresultat}, it remains to asymptotically evaluate the lower bound in
Proposition~\ref{prop:basin} for small energies. This is the topic of the present Section. In order to do so, we
first construct an upper bound on the lowest Dirichlet eigenvalue showing up in the left-hand side of
(\ref{eq:basic}) when choosing
\begin{equation}\label{def:Lambda}
    \Lambda := \overline{\bigcup_{| j |_{} < L/4} \Lambda_j}^{\, \rm int}
\end{equation}
with $ L > 0 $ there. By construction $ \Lambda $ is open and compatible with the lattice.

\subsection{Upper bound on lowest Dirichlet eigenvalue}
The following lemma basically repeats \cite[Prop.~5]{KirSim86} and its corollary.
\begin{lemma}\label{lemma:DN}
  Let $ \Lambda $
  denote the open cube (\ref{def:Lambda}).
  There exist two constant $ 0 < C_1 , C_2 < \infty $
  (which are independent of $ \omega $ and $ L $) such that
  \begin{equation}\label{eq:D1}
    \lambda_0\left( H_\Lambda^D(V_\omega) \right) \leq C_1 \, |\Lambda|^{-1}  \int_\Lambda V_\omega(x) \, dx + C_2 \, L^{-2}
  \end{equation}
  for all $ \omega \in \Omega $ and all $ L  > 1 $.
\end{lemma}
\begin{proof}
 We let $ \theta \in \mathcal{C}_c^\infty(\Lambda_0) $ denote a smoothed indicator function of the cube
 $ \{ x \in \mathbbm{R}^d  :  | x |{} < 1/4 \} \subset \Lambda_0 $ and set $ \theta_L(x) := \theta\big(x/| \Lambda |^{1/d}\big) $
 for all $ x \in \Lambda $.
 Choosing the product of $ \theta_L \in \mathcal{C}_c^\infty(\Lambda)$
 and the ground-state function $ \psi $ of $ H(0) $
 as the variational function in the Rayleigh-Ritz principle we obtain
 \begin{eqnarray}
   && \lambda_0\left( H_\Lambda^D(V_\omega) \right) \, \big\langle \theta_L \psi , \theta_L \psi \big\rangle
   \leq
   \left\langle \theta_L \psi , H_\Lambda^D(V_\omega) \theta_L \psi \right\rangle \nonumber \\
   && \mkern100mu  =  \big\langle \theta_L \psi , V_\omega \theta_L \psi \big\rangle
    + \big\langle \left(\nabla\theta_L\right) \psi ,  \left(\nabla\theta_L\right) \psi \big\rangle \nonumber \\
  && \mkern100mu  \leq \sup_{y \in \Lambda_0 } \psi(y)^2 \; \left[ \int_\Lambda V_\omega(x) \, dx
          +  | \Lambda |^{1-2/d} \int_{\Lambda_0} |\nabla \theta(x) |^2 dx \right].\quad
 \end{eqnarray}
 Here the equality uses $  H_\Lambda^\chi(0) \psi = 0 $ and integration by parts.
 Observing that $ \langle \theta_L \psi , \theta_L \psi \rangle \geq 2^{-d} |\Lambda |\, \inf_{x \in \Lambda_0} \psi(x)^2 $ and
 that the is some constant $ C > 0 $ such that $ | \Lambda |^{1/d} \geq C L $ for all $ L > 1 $, completes the proof.
\end{proof}
Our next task is to bound the integral in the right-hand side of (\ref{eq:D1}) from above. For this purpose it
will be useful to introduce the cuboid
  \begin{equation}\label{def:Lambda2}
    \widetilde \Lambda :=  \bigcup_{\substack{|j_1|_{} \leq 2  L^{\beta_1} \\|j_2|_{} \leq 2 L^{\beta_2} }} \Lambda_j,
  \end{equation}
which contains the cube $ \Lambda $ defined in (\ref{def:Lambda}). Here and in the following we use the
abbreviation $  \beta_k  :=
    \max\left\{ 1 , 2/ \alpha_k (1- \gamma)  \right\} = 2/ d_k \, \max\left\{ d_k/2 , \gamma_k/(1-\gamma)\right\} $, for $ k\in \{1,2\}$.
\begin{lemma}\label{lemma:intV}
  Let  $ L > 0 $ and define the random variable
  \begin{equation}
    W_{\omega}(L) := | \Lambda |^{-1} \int_{\mathbbm{R}^d \backslash \widetilde \Lambda}
  \left(\int_\Lambda f(x-y) \, dx \right)  \mu_{\omega}(dy).
  \end{equation}
  Then the following three assertions hold true:
  \begin{indentnummer}
  \item\label{ident}
    $ | \Lambda |^{-1} \int_\Lambda V_\omega(x) \, dx
    \leq  \left\| f \right\|_1 \, \mu_{\omega}\big( \widetilde \Lambda \big) + W_{\omega}(L) $
    for all $ \omega \in \Omega $ and all $ L > 0 $.
  \item
    there exists some constant $ 0 < C_3 < \infty $ (which is independent of $ \omega $ and $ L $) such that
    \begin{equation}\label{eq:cheb}
      \mathbbm{P}\left\{\omega \in \Omega \, : \,  W_\omega(L) \geq C_3 L^{-2} \right\} \leq \frac{1}{2}
    \end{equation}
    for large enough $ L $.
  \item
    the random variables $ \mu\big( \widetilde \Lambda \big) $ and $ W(L) $ are independent for all $ L > 0 $.
  \end{indentnummer}
\end{lemma}
\begin{proof}
  For a proof of the first assertion we decompose the domain of integration and use Fubini's theorem to obtain
  \begin{eqnarray}\label{eq:add}
     \int_\Lambda V_\omega(x) \, dx
    & = & \int_{\widetilde \Lambda}  \left(\int_\Lambda f(x-y) \, dx \right) \mu_{\omega}(dy)
    + \int_{\mathbbm{R}^d \backslash \widetilde \Lambda} \left(\int_\Lambda f(x-y) \, dx \right) \mu_{\omega}(dy) \notag \\
    & \leq &  \left\| f \right\|_1 \, \mu_{\omega}\big( \widetilde \Lambda \big) + | \Lambda | \, W_{\omega}(L).
  \end{eqnarray}
  Here
  the inequality results from the estimate
  $ \int_{\Lambda} f(x-y) \, dx \leq  \int_{\mathbbm{R}^d} f(x) \, dx =: \| f \|_1 $ valid for all $ y \in \mathbbm{R}^d $.
  This yields Lemma~\ref{ident} since
  $ 1 \leq | \Lambda | $. For a proof of the second assertion, we employ Chebychev's inequality
  \begin{eqnarray}
    \mathbbm{P}\Big\{\omega \in \Omega \, : \,  W_\omega(L) \geq C_3 L^{-2} \Big\}
     & \leq &   \frac{L^2}{C_3} \, | \Lambda|^{-1} \, \mathbbm{E}\left[
     \int_{\mathbbm{R}^d \backslash \widetilde \Lambda}
     \left(\int_\Lambda f(x-y) \, dx \right)  \mu(dy) \right]  \nonumber \\
     & = & \frac{L^2}{C_3} \, | \Lambda|^{-1}
     \int_{\Lambda} \left(\int_{\mathbbm{R}^d \backslash \widetilde \Lambda} f(x-y) \, dx \right)  \overline{\mu}(dy) \nonumber \\
     & \leq & \frac{L^2}{C_3} \, \overline{\mu}(\Lambda_0) \,
     \sup_{y \in \Lambda} \, \int_{\mathbbm{R}^d \backslash \widetilde \Lambda} f(x-y) \, dx.
     \label{eq:twoterms}
  \end{eqnarray}
  Here the inequality uses the fact that the intensity measure $ \overline{\mu} $ is $ \mathbbm{Z}^d $-periodic.
  The inner integral is in turn estimated  from above in term of two integrals involving the marginal impurity potentials $ f^{(1)} $
  and $ f^{(2)} $ (recall the definitions~(\ref{eq:deff2}) and (\ref{eq:deff1}))
  \begin{eqnarray}
    \int_{\mathbbm{R}^d \backslash \widetilde \Lambda} f(x-y) \, dx  & \leq &
    \int_{|x_1|{} > L^{\beta_1}} \mkern-20mu f^{(1)}(x_1-y_1) \, dx_1
    + \int_{|x_2|{} > L^{\beta_2}} \mkern-20mu f^{(2)}(x_2-y_2) \, dx_2 \nonumber \\
    & \leq &  C L^{-2}.
  \end{eqnarray}
  Here the existence of some $ 0 < C < \infty $ ensuring the last inequality for all $ |y|{} \leq L/2 $ (that is in particular; for all $ y \in \Lambda $)
  and sufficiently large $ L \geq 4 $ follows from (\ref{eq:nuetz}) and the fact that
  $  \beta_k\alpha_k(1-\gamma) \leq 2$. Taking $ C_3 $ in (\ref{eq:twoterms}) large enough yields the second assertion.
  The third assertion is a consequence of Assumption~\ref{indep}.
\end{proof}

\subsection{Proof of Theorem~\ref{Thm:Hauptresultat} -- final parts}
For a given energy $ E > 0 $ we choose
\begin{equation}\label{eq:defL}
  L := \left(\frac{3 \max\{C_2,C_3 \}}{E}\right)^{1/2},
\end{equation}
where the constants $C_2$ and $ C_3 $ were fixed in Lemma~\ref{lemma:DN} and Lemma~\ref{lemma:intV},
respectively. Moreover, we pick the cube $ \Lambda $ from (\ref{def:Lambda}) and the cuboid $ \widetilde \Lambda $
from (\ref{def:Lambda2}). Employing Lemma~\ref{lemma:DN} and Lemma~\ref{lemma:intV}  we estimate the probability
in the right-hand side of (\ref{eq:basic}) according to
\begin{align}
  & \mathbbm{P}\left\{\omega \in \Omega \, : \,  \lambda_0\left( H_\Lambda^D(V_\omega) \right) < E \right\}  \nonumber \\
  &\geq \mathbbm{P}\Big(\left\{\omega \in \Omega \, : \,  \lambda_0\left( H_\Lambda^D(V_\omega) \right) < E \right\}
    \cap  \left\{\omega \in \Omega \, : \,  W_\omega(L) < C_3 L^{-2} \right\} \Big) \nonumber \\
  & \geq \mathbbm{P}\Bigg(\Bigg\{\omega \in \Omega \, : \, \mu_{\omega}\big(\widetilde \Lambda\big)
              < \frac{\max\{C_2,C_3 \}\,  L^{-2}}{C_1 \, \| f\|_1 }  \Bigg\}
              \cap  \Big\{\omega \in \Omega \, : \,  W_\omega(L) < C_3 L^{-2} \Big\} \Bigg). \label{eq:mu}
\end{align}
Since the random variables $ \mu( \widetilde \Lambda) $ and  $ W(L)  $ are independent, the probability in
(\ref{eq:mu}) factorises. Thanks to (\ref{eq:cheb}) the probability of the second event is bounded from below by $
1/2 $ provided that $ L $ is large enough, equivalently, that $ E > 0 $ is small enough. Employing the
decomposition (\ref{def:Lambda}) of $ \widetilde \Lambda $ into $ | \widetilde \Lambda | $ unit cubes of the
lattice $ \mathbbm{Z}^d $, we have $ \mu_{\omega}\big(\widetilde \Lambda\big) = \sum_{j \in  \widetilde \Lambda
\cap \mathbbm{Z}^d } \mu_{\omega}\big(\Lambda_j\big) $ such that the probability of the first event in
(\ref{eq:mu}) is bounded from below by
\begin{equation}     \label{eq:unabh}
  \mathbbm{P}\Bigg\{\omega \in \Omega \, : \, \mu_{\omega}\big(\Lambda_j\big)
    <      \frac{\max\{C_2,C_3 \} \, L^{-2}}{C_1 \,  \| f\|_1 \, | \widetilde \Lambda | }  \quad
    \mbox{for all $ j \in \widetilde \Lambda \cap \mathbbm{Z}^d $} \Bigg\}.
\end{equation}
By construction of $  \widetilde \Lambda $ there is some constant $ n_0 > 0 $ such that $ | \widetilde \Lambda |
\leq n_0 \, L^{\beta_1 d_1 + \beta_2 d_2} $. Abbreviating $ C_4 := \max\{C_2,C_3 \} /(C_1 \| f\|_1 \, n_0) $ and $
\vartheta := 2 +\beta_1 d_1 + \beta_2 d_2 $, and using the fact that the random variables $ \mu(\Lambda_j) $ are
independent and identically distributed (by virtue of Assumption~\ref{ass:q}), the last expression
(\ref{eq:unabh}) may be bounded from below by
\begin{eqnarray}
    \mathbbm{P}\Big\{\omega \in \Omega \, :\, \mu_\omega(\Lambda_0) <
             C_4 L^{-\vartheta} \Big\}^{n_0 \, L^{\beta_1 d_1 + \beta_2 d_2} }
     \geq \left( C_4 L^{-\vartheta} \right)^{\kappa n_0 \, L^{\beta_1 d_1 + \beta_2 d_2}} & & \nonumber \\
     = \exp\left[ C_5 \left( \log E^{\vartheta/2} + \log C_6 \right) E^{-(\beta_1 d_1 + \beta_2 d_2)/2} \right]. &&
     \label{eq:finish}
\end{eqnarray}
Here the first inequality derives from Assumption~\ref{ass:q} on the probability measure of $ \mu(\Lambda_0) $.
Moreover, the existence of two constants $ 0 < C_5 $, $C_6 < \infty $ ensuring the validity of the equality
follows from (\ref{eq:defL}). Since the choice (\ref{eq:defL}) of the energy-dependence of $ L $ guarantees that
the pre-factor in the lower bound in Proposition~\ref{prop:basin} is negligible, the proof of
Theorem~\ref{Thm:Hauptresultat} is completed by inserting (\ref{eq:finish}) in the left-hand side of
(\ref{eq:basic}). \qed

\appendix
\section{Proof of mixing of random Borel measure}\label{app:mixing}
The purpose of this short appendix is to proof Lemma~\ref{lemma:mix}. We let $ \Lambda^{(n)} := \bigcup_{| j|{}
\leq n } \Lambda_j $ with $ n \in \mathbbm{N} $. Moreover, let $ \mathcal{M}\big(\Lambda^{(n)}\big) \subset
\mathcal{M}\big(\mathbbm{R}^d\big) $ denote the set of Borel measures with support in $ \Lambda^{(n)} $ and let $
\mathcal{B}(\mathcal{M}_n) $ be the smallest $ \sigma $-algebra, which renders the mappings $
\mathcal{M}\big(\Lambda^{(n)}\big) \ni \nu \mapsto \nu(\Lambda) $ measurable for all Borel sets $ \Lambda \subset
\Lambda^{(n)} $. Their union $ \mathcal{R} := \bigcup_{n \in \mathbbm{N}} \mathcal{B}(\mathcal{M}_n) $ satisfies:
\begin{indentnummer*}
  \item $ \mathcal{R} $ generates the $ \sigma $-algebra $ \mathcal{B}(\mathcal{M}) $.
  \item $ \mathcal{R} $ is a semiring.
\end{indentnummer*}
The first assertion holds by definition of $ \mathcal{B}(\mathcal{M}) $. To check the second one we note that $
\emptyset \in \mathcal{R} $. Moreover, for every $ M $, $ M' \in \mathcal{R} $ there exists some  $ n \in
\mathbbm{N} $ such that
\begin{equation}\label{eq:end}
  M, \, M' \in  \mathcal{B}(\mathcal{M}_n)
\end{equation}
and hence $ M \cap M' \in \mathcal{B}(\mathcal{M}_n) \subset \mathcal{R} $ and  $ M \backslash M' \in
\mathcal{B}(\mathcal{M}_n) \subset \mathcal{R} $.

Our next aim is to prove the claimed limit relation (\ref{eq:mix}) for all $ M $, $ M' \in
\mathcal{B}(\mathcal{M}_n) $ with $ n \in \mathbbm{N} $ arbitrary. Assumption~\ref{indep} ensures that the events
$  T_{j} M \subset \mathcal{M}\big(\Lambda^{(n)}+j\big)  $ and $   M' \subset \mathcal{M}\big(\Lambda^{(n)}\big) $
are stochastically independent for all $ j \in \mathbbm{Z}^d $ with $ \big(\Lambda^{(n)}+j \big)\cap \Lambda^{(n)}
= \emptyset $, such that
\begin{equation}\label{eq:mixp}
  \mathcal{P}\left\{ T_{j} M \cap M' \right\} =  \mathcal{P}\left\{ T_{j} M \right\} \mathcal{P}\left\{M' \right\}
  = \mathcal{P}\left\{ M \right\} \mathcal{P}\left\{M' \right\}.
\end{equation}
Here the last equality is a consequence of Assumption~\ref{stat}.

Thanks to (\ref{eq:end}) we have thus proven the validity of (\ref{eq:mixp}) for all $ M $, $ M' \in \mathcal{R}
$. Lemma~\ref{lemma:mix} now follows from \cite[Lemma 10.3.II]{DaVeJo88}, which is a monotone-class argument. \qed

\begin{remark}
We proved above that the random potential $V_\omega$ is mixing under our assumptions. Note, that mixing is
actually a property of the probability measure $\mathcal{P}$ with respect to the shifts $\left\{T_j\right\}$.
However, the potential $V_\omega$ will not satify stronger mixing condition such as $\phi-$mixing. In fact, as a
rule, the potential may even be deterministic (in the technical sense of this notion, see e.g.\
\cite{KirKotSim85}), which allows mixing, but not $\phi-$mixing. For further references to this see \cite{Bil68,
KirMar83a}.
\end{remark}

\bibliography{bib}

\begin{thebibliography}{BHKL95}

\bibitem[Bil68]{Bil68}
P.~Billingsley.
\newblock {\em Convergence of probability measures}.
\newblock Wiley 1968.

\bibitem[BHKL95]{BHKL95}
K.~Broderix, D.~Hundertmark, W.~Kirsch, and H.~Leschke.
\newblock The fate of {L}ifshits tails in magnetic fields.
\newblock {\em J. Stat. Phys.}, 80:1--22, 1995.

\bibitem[CL90]{CaLa90}
R.~Carmona and J.~Lacroix.
\newblock {\em Spectral theory of random {S}chr\"odinger operators}.
\newblock Birkh\"auser, Boston, 1990.

\bibitem[Dur96]{Dur96}
R.~Durrett.
\newblock {\em Probability: theory and examples}.
\newblock Duxbury, Belmont, 1996.

\bibitem[DV75]{DonVar75}
M.~D. Donsker and S.~R.~S. Varadhan.
\newblock Asymptotics of the {W}iener sausage.
\newblock {\em Commun. Pure Appl. Math.}, 28:525--565, 1975.
\newblock Errata: {\it ibid}, pp.~677.

\bibitem[DVJ88]{DaVeJo88}
D.~J. Daley and D.~Vere-Jones.
\newblock {\em An introduction to the theory of point processes}.
\newblock Springer, New York, 1988.

\bibitem[DZ98]{DeZe98}
A.~Dembo and O.~Zeitouni.
\newblock {\em Large deviations techniques and applications}.
\newblock Springer, New York, 1998.

\bibitem[Erd98]{Erd98}
L.~Erd{\H{o}}s.
\newblock {L}ifschitz tail in a magnetic field: the nonclassical regime.
\newblock {\em Probab. Theory Relat. Fields}, 112:321--371, 1998.

\bibitem[Erd01]{Erd01}
L.~Erd{\H{o}}s.
\newblock {L}ifschitz tail in a magnetic field: coexistence of the classical
  and quantum behavior in the borderline case.
\newblock {\em Probab. Theory Relat. Fields}, 121:219--236, 2001.

\bibitem[HKW03]{HuKiWa03}
D.~Hundertmark, W.~Kirsch, and S.~Warzel.
\newblock {L}ifshits tails in three space dimensions: impurity potentials with
  slow anisotropic decay.
\newblock {\em Markov Process. Relat. Fields}, 9:651--660, 2003.

\bibitem[HLW99]{HuLeWa99}
T.~Hupfer, H.~Leschke, and S.~Warzel.
\newblock Poissonian obstacles with {G}aussian walls discriminate between
  classical and quantum {L}ifshits tailing in magnetic fields.
\newblock {\em J. Stat. Phys.}, 97:725--750, 1999.

\bibitem[HLW00]{HuLeWa00}
T.~Hupfer, H.~Leschke, and S.~Warzel.
\newblock The multiformity of {L}ifshits tails caused by random {L}andau
  {H}amiltonians with repulsive impurity potentials of different decay at
  infinity.
\newblock {\em AMS/IP Stud. Adv. Math.}, 16:233--247, 2000.

\bibitem[Kal83]{Kal83}
O.~Kallenberg.
\newblock {\em Random measures}.
\newblock Akademie-Verlag, Berlin, 1983.

\bibitem[Kir89]{Kir89}
W.~Kirsch.
\newblock Random {S}chr{\"o}dinger operators: a course.
\newblock In H.~Holden and A.~Jensen, editors, {\em {S}chr{\"o}dinger
  Operators}, volume 345 of {\em Lecture notes in physics}, pages 264--370.
  Springer, 1989.

\bibitem[Klo99]{Klo99}
F.~Klopp.
\newblock Internal {L}ifshits tails for random perturbations of periodic
  {S}chr{\"o}dinger operators.
\newblock {\em Duke Math. J.}, 98:335--369, 1999.
\newblock {E}rratum: mp\_arc 00-389.

\bibitem[Klo02]{Klo02}
F.~Klopp.
\newblock Une remarque \'a propos des asymptotiques de {L}ifshitz internes.
\newblock {\em C. R. Acad. Sci. Paris Ser. I}, 335:87--92, 2002.

\bibitem[KKS85]{KirKotSim85}
W.~Kirsch, S.~Kotani, and B.~Simon
\newblock Absence of
  absolutely continuous spectrum for some one-dimensional random but
  deterministic {S}chr\"odinger operators. 
\newblock {\em Ann. Inst. H. Poincare Phys.
  Theor.}, 42:383 -- 406, 1985.


\bibitem[KM82]{KirMar82}
W.~Kirsch and F.~Martinelli.
\newblock On the ergodic properties of the spectrum of general random
  operators.
\newblock {\em J. Reine Angew. Math.}, 334:141--156, 1982.

\bibitem[KM83a]{KirMar83a}
W.~Kirsch and F.~Martinelli.
\newblock Large deviations and {L}ifshitz singularity of the integrated density
  of states of random {H}amiltonians.
\newblock {\em Commun. Math. Phys.}, 89:27--40, 1983.

\bibitem[KM83b]{KirMar83b}
W.~Kirsch and F.~Martinelli.
\newblock On the essential self adjointness of stochastic {S}chr\"{o}dinger
  operators.
\newblock {\em Duke Math. J.}, 50:1255--1260, 1983.

\bibitem[KS86]{KirSim86}
W.~Kirsch and B.~Simon.
\newblock {L}ifshits tails for periodic plus random potentials.
\newblock {\em J. Stat. Phys.}, 42:799--808, 1986.

\bibitem[KS87]{KirSim87}
W.~Kirsch and B.~Simon.
\newblock Comparison theorems for the gap of {S}chr{\"o}dinger operators.
\newblock {\em J. Funct. Anal.}, 75:396--410, 1987.

\bibitem[KW02]{KlWo02}
F.~Klopp and T.~Wolff.
\newblock {L}ifshitz tails for $2$-dimensional random {S}chr{\"o}dinger
  operators.
\newblock {\em J. Anal. Math.}, 88:63--147, 2002.

\bibitem[Lan91]{Lan91}
R.~Lang.
\newblock {\em Spectral theory of random {S}chr{\"o}dinger operators}, volume
  1498 of {\em Lecture notes in mathematics}.
\newblock Springer, Berlin, 1991.

\bibitem[Lif63]{Lif63}
I.~M. Lifshitz.
\newblock Structure of the energy spectrum of the impurity bands in disordered
  solid solutions.
\newblock {\em Sov. Phys. JETP}, 17:1159--1170, 1963.
\newblock Russian original: {\em Zh. Eksp. Ter. Fiz.}, 44:1723--1741, 1963.

\bibitem[LMW03]{LeMuWa03}
H.~Leschke, P.~M{\"u}ller, and S.~Warzel.
\newblock A survey of rigorous results on random {S}chr{\"o}dinger operators
  for amorphous solids.
\newblock {\em Markov Process. Relat. Fields}, 9:729--760, 2003.

\bibitem[LW04]{LeWa03}
H.~Leschke and S.~Warzel.
\newblock Quantum-classical transitions in {L}ifshits tails with magnetic
  fields.
\newblock {\em Phys. Rev. Lett.}, 8:086402 (1-4), 2004.

\bibitem[Mez86]{Mez86}
G.~A. Mezincescu.
\newblock Internal {L}ifshitz singularities for disordered finite-difference
  operators.
\newblock {\em Commun. Math. Phys.}, 103:167--176, 1986.

\bibitem[Mez87]{Mez87}
G.~A. Mezincescu.
\newblock {L}ifschitz singularities for periodic operators plus random
  potential.
\newblock {\em J. Stat. Phys.}, 49:1181--1190, 1987.

\bibitem[Mez93]{Mez93}
G.~A. Mezincescu.
\newblock Internal {L}ifshitz singularities for one dimensional
  {S}chr{\"o}dinger operators.
\newblock {\em Commun. Math. Phys.}, 158:315--325, 1993.

\bibitem[Min02]{Min02}
T.~Mine.
\newblock The uniqueness of the integrated density of states for the
  {S}chr{\"o}dinger operators for the {R}obin boundary conditions.
\newblock {\em Publ. RIMS, Kyoto Univ.}, 38:355--385, 2002.

\bibitem[Nak77]{Nak77}
S.~Nakao.
\newblock On the spectral distribution of the {S}chr{\"o}dinger operator with
  random potential.
\newblock {\em Japan. J. Math.}, 3:111--139, 1977.

\bibitem[Pas77]{Pas77}
L.~A Pastur.
\newblock Behavior of some {W}iener integrals as $t \to \infty$ and the density
  of states of {S}chr\"odinger equations with random potential.
\newblock {\em Theor. Math. Phys.}, 32:615--620, 1977.
\newblock {R}ussian original: {\em Teor. Mat. Fiz.}, 6:88-95, 1977.

\bibitem[PF92]{PaFi92}
L.~Pastur and A.~Figotin.
\newblock {\em Spectra of random and almost-periodic operators}.
\newblock Springer, Berlin, 1992.

\bibitem[RS78]{ReSi4}
M.~Reed and B.~Simon.
\newblock {\em Methods of modern mathematical physics IV: analysis of
  operators}.
\newblock Academic, New York, 1978.

\bibitem[Sim82]{Sim82}
B.~Simon.
\newblock {S}chr{\"o}dinger semigroups.
\newblock {\em Bull. Amer. Math. Soc. (N. S.)}, 7:447--526, 1982.
\newblock Erratum: {Bull. Amer. Math. Soc. (N. S.)}, {1982}, {7}, {447--526}.

\bibitem[Sim85]{Sim85}
B.~Simon.
\newblock {L}ifshitz tails for the {A}nderson model.
\newblock {\em J. Stat. Phys.}, 38:65--76, 1985.

\bibitem[Sim87]{Sim87}
B.~Simon.
\newblock Internal {L}ifshitz tails.
\newblock {\em J. Stat. Phys.}, 46:911--918, 1987.

\bibitem[SKM87]{SKM87}
D.~Stoyan, W.~S. Kendal, and J.~Mecke.
\newblock {\em Stochastic geometry and its applications}.
\newblock Wiley, Chichester, 1987.

\bibitem[Sto99]{Sto99}
P.~Stollmann.
\newblock {L}ifshitz asymptotics via linear coupling of disorder.
\newblock {\em Math. Phys. Anal. Geom.}, 2:2679--289, 1999.

\bibitem[Sto01]{Sto01}
P.~Stollmann.
\newblock {\em Caught by disorder: bound states in random media}.
\newblock Birkh\"auser, Boston, 2001.

\bibitem[Ves03]{Ves03}
I.~Veselic.
\newblock Integrated density of states and {W}egner estimates for random
  {S}chr{\"o}dinger operators.
\newblock {\em preprint math-ph/0307062}, 2003.

\bibitem[War01]{War01}
S.~Warzel.
\newblock {\em On {L}ifshits tails in magnetic fields}.
\newblock Logos, Berlin, 2001.
\newblock {P}hD thesis, University Erlangen-N\"urnberg 2001.

\end{thebibliography}

\end{document}